\documentclass[letter,11pt]{article}

\usepackage[colorlinks=true,linkcolor= red, citecolor=blue]{hyperref}

\usepackage{subfig}

\RequirePackage[OT1]{fontenc}
\advance\textwidth by 2cm

\advance\textheight by 3.8cm

\usepackage{amsmath}
\usepackage{amsfonts,amssymb}
\usepackage{amsthm}
\usepackage{layout}
\usepackage{graphicx,graphics}
\usepackage{multirow}

\usepackage{dsfont}
\usepackage{enumerate}
\usepackage{color}

\numberwithin{equation}{section}

\usepackage{xcolor}

\theoremstyle{plain}
\newtheorem{theorem}{Theorem}[section]

\newtheorem{corollary}[theorem]{Corollary}
\newtheorem{lemma}[theorem]{Lemma}

\newtheorem{Assumption}{Assumption}

\theoremstyle{definition}
\newtheorem{definition}[theorem]{Definition}
\theoremstyle{remark}
\newtheorem{remark}[theorem]{Remark}
\newtheorem{example}[theorem]{Example}

\usepackage{graphicx}

\newcommand\supp{\operatorname{supp}}

\newcommand{\N}{\mathcal{N}}

\newcommand{\calP}{\mathcal{P}}

\newcommand{\calA}{\mathcal{A}}

\def\calX{\mathcal{X}}
\def\calY{\mathcal{Y}}
\def\hpi{\hat{\pi}}

\def\R{{\mathds R}}

\newcommand{\Exp}{\mathds{E}}

\newcommand{\Prob}{\mathds{P}}

\newcommand{\Y}{\mathcal{Y}}

\newcommand{\abs}[1]{\left\vert#1\right\vert}
\newcommand{\set}[1]{\left\{#1\right\}}
\newcommand{\seq}[1]{\left<#1\right>}

\newcommand{\B}{\mathcal{B}}

\usepackage[auth-lg]{authblk}

\title{\textsf{{ALM for insurers with multiple underwriting lines and portfolio constraints: a Lagrangian duality approach}}}

\author[1]{Camilo Castillo\thanks{E-mail: camiloa.castillo@urosario.edu.co}}
\author[1]{Rafael Serrano\thanks{E-mail: rafael.serrano@urosario.edu.co}}

\affil{\smallskip\small \textsc{Universidad del Rosario}\\
	Calle 12C No. 4-69\\
	Bogot\'a, Colombia}

\begin{document}
	

	\date{\today}

	\maketitle




%

\begin{abstract}
	We study a continuous-time asset-allocation problem for an insurance firm that backs up liabilities from multiple non-life business lines with underwriting profits and investment income. The insurance risks are captured via a multidimensional jump-diffusion process with a multivariate compound Poisson process with dependent components, which allows to model claims that occur in different lines simultaneously. Using Lagrangian convex duality techniques, we provide a general verification-type result for investment-underwriting strategies that maximize expected utility from the dividend payout rate and final wealth over a finite-time horizon. We also study the precautionary effect on earnings retention of risk aversion, prudence, portfolio constraints and multivariate insurance risk. We find an explicit characterization of optimal strategies under CRRA preferences. Numerical results for two-dimensional examples with policy limits illustrate  the impact of co-integration for ALM with multiple (dependent and independent) sources of insurance risk. 
\end{abstract}


\section{Introduction}
Insurance is primarily a liability-driven business. Insurers have the responsibility to invest premiums efficiently in order to meet contractual obligations of its existing policies as well as increase wealth and maximize shareholders value. Asset-liability management (ALM) has become the fundamental tool to achieve these goals in the insurance business as it considers the various interrelations between asset classes, underwriting lines, and the time structure of investment cash flows and claim payments.

One of the most prominent strategies for insurance ALM is finding the portfolio with the optimal risk-return trade-off that matches the insurer liabilities. Despite bonds and fixed-income securities dominating portfolios of insurers, their equity allocation has increased considerably for non-life and composite insurers in some countries, possibly because of persistent low interest rates over the past years. Indeed, according to 2019 Global Insurance Market Trends report from the OECD, non-life insurers in Austria, El Salvador, France, Iceland, Poland and Sweden held more than almost 27\% of their assets in equities, and non-life and composite insurers in Argentina, Brazil, Canada, Latvia Germany and Israel invested more than 25\% of their assets in collective investment schemes.

In this work, we consider a theoretical continuous-time portfolio allocation problem for a firm that invests in the financial market and simultaneously holds a portfolio of insurance liabilities in different lines of business. The firm can select both the investments and the volume of underwriting in each business line, with the insurance liabilities being treated as short positions within the overall portfolio. The firm's preferences are represented by a risk-averse utility function, and the goal is to find the investment-insurance strategy that maximizes expected utility from inter-temporal dividend payments and final wealth over a finite-time horizon.

Most existing results in the related literature, except possibly for the work of Zou and Cadenillas \cite{zou2014optimal}, find an optimal investment strategy for a given structure of the insurance portfolio. In contrast, our model allows the structure of the volume of the insurance business to change, thus providing a true ALM framework in which both liability exposure as well as financial risks associated with the investments backing the liability cash flows can be managed simultaneously. The insurance risks arising from the different underwriting lines are captured via a (multidimensional) jump-diffusion process with a multivariate compound Poisson process. This allows to model events that may give rise to claims in different lines simultaneously, for instance, work-related accidents that result in claims for medical and allowance costs, or natural catastrophes that cause damages in homes, vehicles and businesses. Losses caused by wind and water damages, or earthquake and fire damages, can also be highly dependent. This aspect is extremely important in non-life insurance since it has potential implications for pricing, reserving, solvency and capital allocation, see e.g. the book by Denuit et al \cite{denuit2006actuarial}.

The diffusion part of the insurance risk process captures shocks or fluctuations in premiums collected or in the claim values to be processed and paid by the insurance firm. We allow these shocks to be correlated among business lines and with the investments returns, which can in turn be used to model the interdependence between financial assets and insurance liabilities that emerges particularly during recessions, see e.g. Hainaut \cite{hainaut2017contagion}. Indeed, on a short-term basis, rising market volatility leads to a fall in asset prices and deteriorates liquidity, which in turn impact insurance capital, hence higher premium rates, especially for property-casualty (P\&C) insurers.

Declines in interest rates also weigh heavily on the entire insurance industry: lower government bond yields translate into lower discount rates used for the calculation of liabilities, thereby increasing the present value of future payment obligations as well as reinvestment risk, which in turn increases capital requirements. Hence, non-life insurers may reprice insurance contracts in order to mitigate financial risks and make up for eventual losses from their investment portfolios.

Deterioration of economic activity and negative shocks can also lead to a raise in claim payments. In the aftermath of the subprime crisis of 2007-2008, Austria, Luxembourg, Poland, Portugal and Switzerland reported a rise in the range 36\% to 56\% in total gross claim payments, see section titled \emph{Impact of the Financial Turmoil} in the OECD report \cite{OECD2011}. Underwriting lines such as credit insurance were heavily affected during the economic downturn that followed the subprime crisis. This type of insurance offers protection to firms supplying goods and services on credit against nonpayment by their clients. In OECD countries, the implicit or explicit provision of credit by sellers to buyers was common practice in the years preceding the crisis. Countries like Spain, France and the U.K, used credit insurance to cover over EUR 200 billion, EUR 320 billion and \textsterling 300 billion respectively, according to the OECD report \cite{OECD2011}. After the crisis, Europe credit insurers increased on average their premiums up to 30\% for renewal business and up to 60\% for new business, see Marovi{\'c} et al \cite{marovic2010implications}.

The early stages of the recession caused by the recent COVID-19 pandemic outbreak also impacted heavily both financial and insurance sectors. Underwriting lines such as travel insurance, short-term disability, business interruption and other specialty lines faced mounting claims in the wake of the outbreak. In April 2020, the property-casualty industry estimated that business interruption losses from small businesses in the U.S. due to the COVID-19 outbreak could be between \$220 an \$383 billion per month, or a quarter to half of the total industry surplus available to pay all P\&C claims. Conversely, other lines such as car insurance experienced a decline in claims. Personal vehicle travel in the U.S. dropped nearly 50\% due to COVID-19 restrictions, compared with typical traffic volume. Because of fewer car accidents being registered, the insurance industry could end up saving \$100 billion from claims, which should translate to lower premiums for consumers. These examples illustrate how negative economic shocks, especially catastrophic events, make the insurance industry vulnerable to simultaneous shocks in their risk-absorbing capital, which clearly challenges the investment assumption, especially in property and casualty underwriting lines, that there is no major relation between underwriting and investment risks, see also the discussions in Achleitner et al. \cite{achleitner2002}, Baluch et at. \cite{baluch2011insurance}, Baranoff and Sager \cite{baranoff2011interplay},  Ko{\v{c}}ovi{\'c} et al. \cite{kovcovic2011impact} and Schich \cite{schich2010insurance}.

Our approach to the utility maximization problem follows closely the (Lagrangian) convex duality method started by He and Pearson \cite{hepearson}, Karatzas, Lehoczky and Shreve \cite{karatzas91} and Cvitaniv and Karatzas \cite{cvikar} (see also the books by Karatzas and Shreve \cite{karatzas1998methods}) that consists in formulating an associated dual minimization problem and finding conditions for absence of duality gap. This method has been remarkably effective to solve the investment-consumption problem in a jump-diffusion setting (see e.g. Goll and Kallsen \cite{goll2000}, Kallsen \cite{kallsen2000}, Callegaro and Vargiolu \cite{callevar} and Michelbrink and Le \cite{michelbrink}) as well as the investment problem for insurers, see for instance the work by Wang, Xia and Zhang \cite{wang2007optimal} that uses the martingale method with CARA and mean-variance preferences and a Levy-type risk process, Zhou \cite{zhou2009optimal} that obtains closed-form solutions in a similar model with CARA-type utility, and Liu \cite{liu2010optimal} that also uses the martingale method for both CARA and mean-variance preferences and characterizes the mean-variance frontier. More recently, Zou and Cadenillas \cite{zou2014optimal} consider CARA, CRRA and mean-variance preferences, and use the volume of underwriting as a control variable. However, they consider only non-random valued claims.

Rising lending costs and regulatory restrictions  have tightened portfolio allocation constraints in the property and casualty insurance industry. Recently, Reddic \cite{reddic2021under} showed that investment limitations imposed by insurance regulators can inhibit desired investment allocation post the financial crisis. The main theoretical contribution of the present paper is to adapt  successfully  the martingale and convex duality approach to address the optimal ALM problem with random liabilities and portfolio constraints: we prove a general verification-type theorem that provides a sufficient condition for existence of an optimal strategy in terms of the solution of a backward jump-diffusion SDE, and characterize the precautionary effect of risk-aversion and prudence of insurer preferences, as well as portfolio constraints, arrival rates and first-order stochastic dominance of claim distributions, on the earnings retention policy of the firm.

It is worth mentioning that the precautionary earnings analysis does not depend on a particular form of the utility functions, only on its risk aversion and prudence index, which is a clear advantage of our approach over dynamic programming methods that rely on solutions to HJB equations. Finally, we present an explicit characterization of optimal strategies for CRRA power utility preferences with unconstrained portfolios as well as rectangular constraints. This shows that both financial and multivariate underwriting risks can be  hedged partially in an efficient manner in the face of portfolio constraints and extreme events. Numerical examples illustrate that co-integration is important to investment-insurance ALM with multiple (dependent and independent) sources of insurance risk.

Let us briefly describe the contents of this paper. In Section 2 we formulate the models for the financial market and multivariate insurance risk processes, and define the wealth process. In Section 3 we use the martingale method and convex duality approach to solve the optimization problem and formulate the verification theorem. We also see how prudence and the model parameters, particularly the claim arrival rate, impact the growth rate of the optimal retention policy. In Section 4 we focus on the case of an insurance firm with CRRA preferences and obtain semi-closed form solutions in this setting. We also provide numerical examples for bivariate claim distributions with dependence modeled via a Levy copula and policy limits. Section 5 outlines some conclusions of our work.

\section{Market model  and risk-averse ALM problem}
We consider a firm that at time $t=0$ starts underwriting insurance policies and, at the same time, allocates an initial endowment $x>0$ among assets in the financial market. Subsequently, at each time $t>0$ the firm collects insurance premiums,  process and pays insurance claims filed by policyholders, and rebalances allocations in the investment portfolio. The firm also uses part of its wealth to pay dividends to stockholders.

Our setting for the insurance portfolio follows an approach similar to B{\"a}uerle and Blatter \cite{bauerle2011optimal}. We assume the firm underwrites $M\in\mathds N$ different types of insurance and that the risk reserves of each line  $j=1,\ldots,M$ can be managed via the reinsurance/underwriting control variable $L^j\geq 0$ which is used by the firm as a decision variable  as follows: if $L^j< 1$ then $L^j$ is set as the proportional reinsurance retention level, that is, the fraction of the incoming claims that  the firm insures itself. If instead $L^j\geq 1$ then the firm must adjust its underwriting volume (number of policies) so that it reaches the level $L^j.$ In practice the quantity $L^j$ would be an integer number. However, for simplicity we assume all business lines are ``infinitely divisible" so that non-integer numbers are allowed.

We denote with $X^j$ the insurance liability risk process (potential loss per unit of exposure) and $p^j$ the premium rate for business line $j=1,\dots,M.$ We assume the reinsurance/underwriting control variable $L\in[0,\infty)^M$ can also change over time. Then, the dynamics of the P\&L for the insurance portfolio follows the process
\[
G_t^L:=\int_0^tL_\tau\cdot(p_\tau\,d\tau-dX_\tau)=\sum_{i=0}^M \int_0^tL_\tau^j(p_\tau^j\,d\tau-dX_\tau^j), \ \ t\geq 0.
\]
The firm backs the reserves for the insurance liabilities with the premiums received and the returns from investing in a financial market model consisting of one money market account with price process $S^0$ and $K$ non-dividend-paying risky assets or stocks with price-per-share processes $S^i, \ i=1,\ldots,K.$ We denote with $\alpha^i_t$ the number of units of risky asset $S_t^i$  held by the firm at time $t\geq 0.$ Then the value of the holdings in the financial market is
\[
V_t^{\alpha}:=\alpha_t\cdot S_t=\sum_{i=0}^K \alpha_t^iS_t^i, \ \ \ t\geq 0.
\]
Each trading strategy $\alpha_t=(\alpha_t^0,\alpha_t^1,\ldots,\alpha^K_t)$ is associated with a P\&L process defined by
\[
G_t^{\alpha}:=\int_0^t \alpha_\tau\cdot dS_\tau=\sum_{i=0}^K \int_0^t\alpha_\tau^i dS_\tau^i, \ \ t\geq 0.
\]
Throughout, we consider a fixed finite investment interval $[0,T].$ Given an initial endowment $x\in\R,$ the strategy $(\alpha,L)$ is said to be \emph{self-financed} if $V_0^\alpha=x$ and existing resources, right before jumps in $G^L,$ are sufficient to subsidize the investment portfolio $V^\alpha$ over the time interval $[0,T],$  that is, if the following budget constraint holds
\[
V_t^{\alpha}\le x+G_t^{\alpha}+G_{t-}^L, \ \ t\in [0,T].
\]
Earnings that are not reinvested in the financial market or used to pay claims, are paid as dividends to stockholders. More precisely, for a self-financing strategy $(\alpha,L)$ we define the cumulative dividend process
\begin{equation}\label{cum-cons}
{C^{\alpha,L}_t:=x+G_t^{\alpha}+G_{t-}^L-V_t^{\alpha}}, \ \ t\in [0,T].
\end{equation}
As measure of the performance of $(\alpha,L),$ we follow Hubalek and Schachermayer \cite{hubalek2004optimizing} or Liang and Palmowski \cite{liang2018note}, and consider the dividend payouts that can be achieved over the time interval $[0,T].$  More concretely, we say that a self-financing strategy $(\alpha,L)$ is \emph{admissible} if $V_t^{\alpha}>0$ for all $t\in [0,T]$ and the map $[0,T]\ni t\mapsto C^{\alpha,L}_t$ is increasing and absolutely  continuous with respect to the Lebesgue measure. Although the measurement of a utility of a density may seem strange at a first glance, this can be
motivated by interpreting the problem as a limit of a discrete model, where the cumulated utility of the payments from each time step is considered, see e.g. Borch \cite{borch1974mathematical}.

In such case, we define the instantaneous \emph{dividend payout rate} $D^{\alpha,L}$ as the density process $D_t^{\alpha,L}:=dC^{\alpha,L}/dt$ modeling the rate of dividend payments. Using this definition, the equality (\ref{cum-cons}) can be rewritten in differential form as follows
\begin{equation}\label{wealth0}
dV_t^{\alpha}=\alpha_t\cdot dS_t+L_t\cdot(p_t\,dt-dX_t)-D_t^{\alpha,L}\,dt, \ \ V_0^{\alpha}=x.
\end{equation}
We now introduce the stochastic setting for our model. The price processes $S=(S^1,\ldots,S^K)^\top$ follow a Black-Scholes model of the form
\begin{align*}
	dS_t^i&=S_t^i\Bigl[\mu_t^i\,dt+\sum_{k=1}^K\sigma_t^{ik}\,dW_t^k\Bigr], \  S_0^i>0, \ \ i=1,\ldots,K\\
	dS_t^0&=S_t^0r_t\,dt, \ S_0^0=1
\end{align*}
where $W=(W^1,\ldots,W^K)^\top$ is a $K$-dimensional Brownian motion defined on a complete probability space $(\Omega,\Prob,\mathcal{F})$ endowed with a filtration $\mathds F=\set{\mathcal{F}_t}_{t\geq 0}.$ For the multivariate insurance risk process $X_t$, we assume claims in the $M$ business lines can occur simultaneously, and model arrival times and claim severities via a $\R_+^M$-valued marked point process $(\tau_n,Y_n)_{n\geq 1}.$ In order to motivate the specification of $X_t$ in our model, for each $j=1,\ldots,M$ we set $\tau_0^j:=\tau_0=0,$
\[
\tau_n^j:=\inf\bigl\{\tau_k>\tau_{n-1}^j:Y_k^j>0\bigr\}, \ \ n\in\mathds N
\]
and suppose, for the sake of argument, that the local characteristics $(\lambda^j,F^j)$ of $(\tau_n^j,Y_n^j)_{n\geq 1}$ and $L^j$ are known and constant in time. We also assume $L^j\in\mathds N,$ that is, the firm must adjust the underwriting volume of line $j$ so that it reaches the number of contracts $L^j$. Then, the total payout up to time $t$ in line $j$ is a sum of compound Poisson processes of the form
\begin{equation}\label{total-claims}
	\sum_{l=1}^{L^j}\sum_{n=1}^{N_t^{j,l}}Y_n^{j,l}
\end{equation}
where $N_t^{j,l}\sim\mathrm{Poisson}(\lambda^jt)$ and $Y_n^{j,l}\sim F^j.$ 	If the compound Poisson sums $\sum_{n=1}^{N_t^{j,l}}Y_n^{j,l}$ are independent, then (\ref{total-claims}) has the same distribution as $\sum_{n=1}^{N_t^{j}}Y_n^{j}$
with $N_t^{j}\sim\mathrm{Poisson}(L^j\lambda^jt),$ see e.g. Proposition 3.3.4 in Mikosch \cite{mikosch2004}, which in turn has the same compensator of the process
\begin{equation}\label{AL-constant}
	\sum_{\tau_n^j\le t}L^jY_n^j.
\end{equation}
In particular, (\ref{total-claims}) and (\ref{AL-constant})  represent the same expected loss. In our model, we allow $L^j$ to take any non-negative real value and change over time, we incorporate changes in the payout process by interpreting (\ref{AL-constant}) as a (discrete) integral of $L^j$ with respect to the multivariate process $\sum_{\tau_n\le t}Y_n^j$, and consider a perturbed version of this process as proxy for the liabilities from line $j$ in our model. More concretely, we assume that for each $j=1,\ldots,M,$ the liability risk process $X^j$ for business line $j$ follows the jump-diffusion process
\[
X_t^j=\sum_{m=1}^M\int_0^tb_s^{jm}\,d\bar W_s^m+\sum_{\tau_n\le t} Y_n^j
\]
where $\bar W$ is a $M$-dimensional Brownian motion satisfying  $d\seq{W^k,\bar W^m}_t=\rho_t^{km}\,dt$ with {$\rho^{km}_t\in[-1,1].$} All coefficients $r_t,\mu_t,b_t^{jm},\sigma_t^{ik}$ and $\rho_t^{km}$ are locally bounded $\mathds F$-predictable processes, and $(\tau_n,Y_n)_{n\geq 1}$ is assumed independent of $W$ and $\bar W.$ The correlation processes $\rho^{km}$ model the dependence between the log-prices of the financial assets and the (Gaussian) fluctuations in the premiums or in the values of the claims.

Recall that a real-valued process $(\phi_t)_{t\geq 0}$ is $\mathds{F}$-predictable if the random function $\phi(t,\omega)=\phi_t(\omega)$ is measurable with respect to the $\sigma-$algebra $\calP$ on $\Omega\times[0,\infty)$ generated by all adapted left-continuous processes. Similarly, a random field $\phi:\Omega\times[0,\infty)\times \R^M\to\R$ is said to be  $\mathds{F}$-predictable if it is measurable with respect to the product $\sigma$-algebra $\calP\otimes\B(\R^M).$

\begin{remark}
Note that (\ref{AL-constant}) also corresponds to the worst-case scenario in which all customers in line $j$ report claims of severity $Y_m^j$ simultaneously,  with the same claim arrival rates $\lambda^j.$ This can be used to model catastrophic events or negative economic shocks that cause sudden surges in claims for underwriting lines with significant exposure to disaster or extreme-event risk. Credit insurance in the aftermath of the subprime crisis is a clear example of this phenomenon. Indeed, according to the OECD report \cite{OECD2011}, the total annual premium income for credit insurance in 2008 was over USD 8 billion, with 90\% of business conducted by three major firms: Euler Hermes (36\%), Atradius (31\%), and Coface (20\%). Once credit conditions worsened in 2008 and early 2009, credit insurers started facing fast-rising claims as the number of payment defaults and corporate insolvencies soared, with loss ratios rising to 73\% at Coface, 78\% at Euler Hermes, and 99\% at Atradius in 2008. These negative trends continued in the first half of 2009 as Euler and Coface reported  loss ratios of 88\% and 116\% respectively.
	
	The recent COVID-19 pandemic outbreak is another example of an extreme event that caused a rush of insurance claims. The American Property-Casualty Insurance Association (APCIA) anticipated in March 2020 that there could be as many as 30 million claims from small business that suffered coronavirus-related losses, triggering claim payments in the range of USD 220-383 billion in only one month, which is 10 times the most claims ever handled by the industry in one year.
	
	
	Unemployment benefit schemes, which are treated by law as a type of insurance in the U.S., also experienced a sudden surge of claims during the COVID-19 outbreak. This a type of insurance in which employees are beneficiaries and employers pay the premiums via unemployment taxes based on their history of layoffs. In March 2020, the number of Americans who applied for unemployment benefits rocketed to record numbers as large parts of the U.S. economy shut down and companies laid off scores of workers to cope with the pandemic.
\end{remark}
Continuing with the formulation of the ALM problem, as it is usually the case with portfolio allocation problems, we work with the proportion of wealth invested in the risky assets
\[
\pi_t^{\alpha,i}:=\frac{\alpha_t^i S_t^i}{V^{\alpha}_{t-}}, \ \ i=1,\ldots K
\]
instead of $\alpha_t.$ We refer to $\pi_t^\alpha$ as \emph{portfolio proportions} process. Similarly, we define an additional \emph{insurance risk} control variable as follows
\[
\kappa_t^j:=\frac{L_t^j}{V^{\alpha}_{t-}}, \ \ j=,\ldots,M.
\]
This is referred to as `liability ratio' by Cadenillas and Zou \cite{zou2014optimal}. Notice that its reciprocal $1/\kappa_t^j=V_{t-}^\alpha/L^j_t$ is the amount of wealth that backs up the liabilities of each insurance contract in the underwriting line $j=1,\ldots,M.$ It is also related to the investment-income ratio
\[
\frac{1}{p_t^j\kappa_t^j}=\frac{V_{t-}^\alpha}{p_tL_t}
\]
which in turn can be interpreted as a measure of the profitability of business line $j,$ as it compares the income the insurance company brings from underwriting type $j$ insurance policies with its investment activities in the financial market, see e.g. the books by R. Kumar \cite[Section 8.2.7.5]{kumar2014} or \cite[Section 10.2.2]{kumar2015}.


Using $\pi$ and $\kappa$ as control variables instead of $\alpha$ and $L$, equation (\ref{wealth0}) for the firm's reserve process now turns into the linear SDE
\begin{equation}\label{eqVnon-linear0}
dV_t=V_{t-}\biggl\{r_t\,dt+\pi_t^\top[(\mu_t-r_t\underline 1)\,dt+\sigma_t\,dW_t]+\kappa_t^\top\Bigl[p_t\,dt-b_t\,d\bar W_t-y\star N(dy,dt)\Bigr]\biggr\}-D_t\,dt
\end{equation}
with initial condition $V_0=x.$ Here $N(dy,dt)$ denotes the random counting measure on $\R_+^M\setminus\set{\underline 0}$ of the multivariate marked point process $(\tau_n,Y_n)_{n\geq 1}$ and $\star$ denotes componentwise integration with respect to random measures. The firm's reserve is now defined as the process $V^{\pi,\kappa,D}$ solution to equation (\ref{eqVnon-linear0}).

Finally, we assume that the portfolio process $\pi$ is constrained to take values in a set-valued predictable process $Q=(Q_t)_{t\in[0,T]},$ where each $Q_t(\omega)$ is a nonempty, closed, convex set $Q_t(\omega)\subset \R^K.$ We formulate the risk-averse ALM problem for the insurance firm as follows: let $U_1(t,\cdot)$ and $U_2$ be utility functions satisfying the usual Inada conditions. We fix throughout the initial wealth $x>0$ and denote with $\mathcal{A}:=\mathcal A(x)$ the set of admissible strategies $(\pi,\kappa,D)$ for which
\[
\Exp\left[\int_0^T U_1(t,D_t)^-\,dt+U_2\left(V_T^{\pi,\kappa,D}\right)^-\right]<\infty.
\]
Our goal is to maximize the expected utility functional
\[
J(x;\pi,\kappa,D):=\Exp\left[\int_0^T U_1(t,D_t)\,dt+U_2\left(V_T^{\pi,\kappa,D}\right)\right]
\]
over all admissible strategies $(\pi,\kappa,D)\in\mathcal A.$ The optimal reinsurance/underwriting strategy can be recovered by defining $\hat L_t^j:=\hat\kappa_t^j \hat V_{t-}$ where $\hat V$ is the optimal value of the total reserve. If $\hat L_t^j<1,$ that is, if $\hat V_{t-}<1/\kappa_t^j$ then the firm must have the fraction $1-\hat L_t^j=1-\hat \kappa_t^j\hat V_{t-}$ of its incoming claims reinsured by a third party. On the contrary, if $\hat L_t^j\geq 1,$ that is, if $\hat V_{t-}\geq 1/\hat \kappa_t^j$ then the firm must readjust the underwriting volume of line $j$ so that it reaches $\hat \kappa_t^j \hat V_{t-}.$

This analysis also provides an interpretation of $1/\hat \kappa_t^j$ as a threshold solvency value. Indeed, if the total value of the reserve is large enough so that $\hat V_{t-}\geq 1/\hat \kappa_t^j,$ the firm can continue underwriting insurance in business line $j,$ although it may need to adjust the number of policies. If instead $\hat V_{t-}< 1/\hat \kappa_t^j,$ the insurer does not hold enough capital against the risks it faces in line $j$ and must transfer part of this risk to a reinsurer.

\begin{remark}\label{DepVsIndep-claims}
Equation (\ref{eqVnon-linear0}) is inhomogeneous linear, so it can be easily solved using variation of parameters and integrating factor. In particular, its solution satisfies $V_t^{\pi,\kappa,D}>0$ if
	\begin{equation}\label{cons-Vpositive}
	\delta_t^{\pi,\kappa,D}:=\int_0^t\frac{D_s}{V_{s-}^{\pi,\kappa,0}}\,ds\le x
	\end{equation}
	almost surely, and $\kappa_{\tau_m}\cdot Y_m<1,$ that is, $V_{\tau_m-}^{\pi,\kappa,D}>L_{\tau_m}\cdot Y_m,$ for $\tau_m\le t.$ The latter is equivalent to requiring the total reserve process to be strictly larger than the aggregated loss from all business lines right before claims occur. This seems too restrictive and impractical but does make sense for business lines with exposure to catastrophic events and high tail dependence, that is, lines with claim  dependence that concentrate in the extreme high values. A simple example of tail dependence comes from wind and water damages. In the U.S. these damages are insured separately: the former is covered under homeowners’ policies or state wind pools, while the latter is covered by the National Flood Insurance Program. Flood and wind damage are often independent but can clearly become tail dependent in hurricane-prone regions.
	
	Another example comes from considering the damage distributions associated with computer networks and highly infectious diseases as during the recent COVID-19 pandemic outbreak. Events in the tail of the damage distribution associated with potential computer network problems include network failure and malicious attacks. Events in the extreme tail of the infectious disease includes not only rising infection and mortality rates, but also mass lockdowns. These negative outcomes, however, are not independent. If people were quarantined at home, the number of people telecommuting would increase dramatically, stressing computer networks and leading to failures and vulnerabilities that could be exploited.

	
	On the other hand, if claims from two different business lines can not occur simultaneously, the above condition can be weakened to $\kappa_{\tau_m^j}^j\cdot Y_m^j<1,$ that is
	\[
	V_{\tau_m^j-}^{\pi,\kappa,D}>L_{\tau_m^j}^j\cdot Y_m^j, \ \mbox{ for } \tau_m^j\le t, \ \mbox{ for all } \ j=1,\ldots,M.
	\]
	This is the same as the wealth process being strictly larger than the loss in each business line right before claims occur, which is much more reasonable from a practical point of view. Note that there is still dependence among underwriting lines through the diffusion part of the multivariate insurance risk process.
\end{remark}

\section{Lagrangian semi-martingale and convex duality approach}
This section extends the convex duality techniques for portfolio constraints from Karatzas and Shreve \cite[Chapter 6]{karatzas1998methods} to the investment-insurance setting. For each $t\in[0,T]$ we define the support function $\vartheta_t$ of the convex set $-Q_t$ as
\[
\vartheta_t(\omega,\zeta):=\sup_{\pi\in Q_t(\omega)}[-\pi\cdot\zeta], \ \ \zeta\in\R^K.
\]
It is a lower semicontinuous, proper (i.e., not identically $+\infty$) convex function, which is finite on its effective domain $\N_t(\omega):=\set{\zeta\in\R^K:\vartheta_t(\omega,\zeta)<+\infty}.$ The latter is a convex cone, called the barrier cone of $-Q_t(\omega).$ In what follows it will be assumed that $\vartheta_t(\omega)$ is bounded from below.

\begin{example}
The following are some examples of possible constraint sets on portfolio proportions.
\begin{itemize}
	\item[i.] Incomplete market: $Q_t=\set{\pi\in\R^K:\pi^i=0, \ i={m+1},\ldots,K}$ for some $m\in\set{1,\ldots,K-1}.$ That is, the firm can only invest in the first $m$ assets. Then $\N_t=\set{\zeta\in\R^K:\zeta^1=\cdots=\zeta^m=0}$ and $\vartheta_t\equiv 0$ on $\N_t.$
	
	\item[ii.] More generally, $Q_t$ is a nonempty, closed, convex cone in $\R^K.$ Then $\N_t$ is the polar cone of $-Q_t$ and  $\vartheta_t\equiv 0$ on $\N_t.$ This includes the case of incomplete markets with and without prohibition of short selling.
	
	\item[iii.] Rectangular constraints: $Q_t=\prod_{k=1}^{K}I_t^k$ with $I_t^k=[\underline q_t^k,{\overline q}_t^k],$ with $\underline q$ and $\overline q$ predictable processes satisfying $-\infty\le\underline q^k\le 0\le\overline q^k\le\infty.$ Here we assume the convention that $I_t^k$ is open on the right (resp. left) if $\overline q_t^k=\infty$ (resp. $\underline q_t^k=-\infty$). Then $\mathcal N_t=\R^K$ and
	\[
	\vartheta_t(\zeta)=\sum_{k=1}^d\overline q_t^k(\zeta^k)^-\underline q_t^k(\zeta^k)^+
	\]
	if all the $\underline q_t^k$ and ${\overline q}_t^k$ are finite. More generally,
	\[
	\mathcal N_t=\set{\zeta\in\R^K:\zeta^i\geq 0 \mbox{ if } \overline q_t^i=\infty, \ \zeta^k\le 0 \mbox{ if } \underline q_t^k=-\infty, \mbox{ for some }i,k=1,\ldots,K}
	\]
	and the previous formula for $\vartheta_t(\zeta)$ remains valid. This includes borrowing and/or short-selling constraints.
\end{itemize}
\end{example}

Let $\mathcal{D}$ the set of $\R^d$-valued predictable processes $\zeta$ satisfying
\begin{equation}\label{calD}
\sup_{t\in [0,T]}\abs{\zeta_t}+\int_0^T\vartheta_t(\zeta_t)\,dt<+\infty, \ \ \mbox{a.s.}
\end{equation}
\begin{Assumption}\label{depend-assumption}
	The multivariate marked point process $(\tau_n,Y_n)_{m\geq 1}$ on $[0,\infty)^M$ has predictable characteristics $(\lambda_t,F_t).$
\end{Assumption}
In what follows we denote $Y_0:=0$ and $Y_t:=Y_n$ for $t\in(\tau_{n-1},\tau_n].$ Let $\Theta$ denote the set of locally bounded pairs $(\theta,\varphi)$ satisfying
\begin{itemize}
	\item[i)] $\theta_t=(\theta_t^1,\theta_t^2)$ is a predictable process with values in $\R^d\times\R^M,$
	\item[ii)] $\varphi=\varphi(t,y)$ is a (real-valued) positive predictable field on $[0,T]\times\R^M,$
\end{itemize}
such that the process
\[
\zeta_t^{\theta}:=r_t\underline 1-\mu_t+\sigma_t[\theta^1_t+\rho_t\theta^2_t], \ \ t\in[0,T]
\]
belongs to $\mathcal D$ and the following condition holds a.s.
\begin{equation}\label{SDFcond1}
p_t+b_t[\rho_t^\top\theta_t^1+\theta^2_t]-\lambda_t\Exp[\varphi(t,Y_t)Y_t]=\underline 0
\end{equation}
for almost every $t\in [0,T].$ The expected value in (\ref{SDFcond1}) is multivalued as it is calculated componentwise. For $(\theta,\varphi)\in\Theta,$ let $H^{\theta,\varphi}$ be the solution of the linear SDE
\[
dH_t=H_{t-}\left\{-[r_t+\vartheta_t(\zeta_t^\theta)]\,dt-\theta_t^1\cdot dW_t-\theta_t^2\cdot\,d\bar W_t+\left[\varphi(t,y)-1\right]\star\widetilde{N}(dy,dt)\right\}
\]
with $H_0=1,$ where $\widetilde{N}$ is the compensated measure $\widetilde{N}(dy,dt):=N(dy,dt)-\lambda_tF_t(dy)\,dt.$ Then, the following deflator-type inequality holds
\begin{lemma}\label{deflator}
	Let $(\theta,\varphi)\in\Theta$ and suppose $V_s^{\pi,\kappa,D}>0$ a.s. for almost every $s\in[0,t].$ Then
	\[
	\Exp\left[H_t^{\theta,\varphi} V_t^{\pi,\kappa,D}+\int_0^t H^{\theta,\varphi}_s D_s\,ds\right]\le x.
	\]
\end{lemma}
\begin{proof}
	See Appendix.
\end{proof}
We refer to $H^{\theta,\varphi}$ as the \emph{Lagrangian semimartingale} for the insurance-investment market model. For a positive random variable $G$ and dividend payout rate process $D,$ we define
\[
\bar J(G,D):=\Exp\Bigl[\int_0^T\!U_1(t,D_t)\,dt+U_2(G)\Bigr]
\]
and
\[
\Lambda^{\theta,\varphi}(G,D):=\Exp\Bigl[H_T^{\theta,\varphi} G+\int_0^T\!H_t^{\theta,\varphi}D_t\,dt\Bigr], \ \ ({\theta,\varphi})\in\Theta.
\]
Then, by Lemma \ref{deflator}, we have
\begin{align*}
\sup_{(\pi,\kappa,D)\in \calA(x)} \,&J(x;\pi,\kappa,D)\\
&\le \sup\Bigl\{\bar J(G,D): G\geq 0, \ D\geq 0, \  \Lambda^{\theta,\varphi}(G,D)\le x, \  \forall ({\theta,\varphi})\in\Theta\Bigr\}
\end{align*}
This suggests to consider the following Lagrangian
\[
L(G,D;\theta,\varphi,\xi):=\bar J(G,D)+\xi\Bigl[x-\Lambda^{\theta,\varphi}(G,D)\Bigr], \ ({\theta,\varphi})\in\Theta, \ y\geq 0.
\]
Then, the following weak duality holds
\begin{align*}
&\sup\Bigl\{\bar J(G,D): G\geq 0, \ D\geq 0, \  \Lambda^{\theta,\varphi}(G,D)\le x, \  \forall(\theta,\varphi)\in\Theta\Bigr\}\\
&=\sup_{\substack{G\geq 0\\ D \geq 0}}\inf_{\substack{({\theta,\varphi})\in\Theta\\\xi\geq 0}}L(G,D;\theta,\varphi,\xi) \ \ \\
&\le\inf_{\substack{({\theta,\varphi})\in\Theta\\\xi\geq 0}}\sup_{\substack{G\geq 0\\ D \geq 0}}L(G,D;\theta,\varphi,\xi)
\end{align*}
Let $U$ denote either $U_2(\cdot)$ or $U_1(t,\cdot)$ with $t\in[0,T]$ fixed. The inverse marginal utility $I:=(U')^{-1}$ satisfies the Young-type inequality $U(x)-\xi x\le U(I(\xi))-\xi I(\xi)$ for all $x,\xi>0.$ Then $L(G,D;{\theta,\varphi},\xi)\le L(I_2(\xi H_T^{\theta,\varphi}),I_1(\cdot,\xi H_\cdot^{\theta,\varphi});{\theta,\varphi},\xi)$ and
\[
\inf_{\substack{{\theta,\varphi}\in\Theta\\\xi\geq 0}}\sup_{\substack{G\geq 0\\D\geq 0}}L(G,D;\varphi,\xi)\le\inf_{\substack{{\theta,\varphi}\in\Theta\\\xi\geq 0}}L(I_2(\xi H_T^{\theta,\varphi}),I_1(\cdot,\xi H_\cdot^{\theta,\varphi});{\theta,\varphi},\xi)
\]
Moreover, it can be shown (see e.g. Lemma 6.2 in Karatzas and Shreve \cite{karatzas1998methods}) that if 
\[
\calX^{\theta,\varphi}(\xi):=\Lambda^{\theta,\varphi}(I_2(\xi H_T^{\theta,\varphi}),I_1(\cdot,\xi H_\cdot^{\theta,\varphi}))<+\infty, \ \ \mbox{ for all } y\geq 0
\]
then its inverse $\calY^{\theta,\varphi}:=(\calX^{\theta,\varphi})^{-1}$ exists and
\[
L(G^{x,{\theta,\varphi}},D^{x,{\theta,\varphi}};{\theta,\varphi},\calY^{\theta,\varphi}(x))=\bar J(G^{x,{\theta,\varphi}},D^{x,{\theta,\varphi}})
\]
with
\begin{align*}
D_t^{x,{\theta,\varphi}}&:=I_1(t,\Y^{\theta,\varphi}(x)H_t^{\theta,\varphi}), \ \ t\in [0,T]\\
G^{x,{\theta,\varphi}}&:=I_2(\Y^{\theta,\varphi}(x)H_T^{\theta,\varphi}).
\end{align*}
In summary, we have
\begin{align*}
\sup_{(\pi,\kappa,D)\in\mathcal A(x)}J(x;\pi,\kappa,D)&\le\sup_{G\geq 0}\inf_{\substack{({\theta,\varphi})\in\Theta\\\xi\geq 0}}L(G,D;{\theta,\varphi},\xi) \ \ \mbox{({Primal})}\\
&\le\inf_{\substack{({\theta,\varphi})\in\Theta\\\xi\geq 0}}\sup_{G\geq 0}L(G,D;{\theta,\varphi},\xi) \\
&\le \inf_{\substack{({\theta,\varphi})\in\Theta\\\xi\geq 0}}L(I_2(\xi H_T^{\theta,\varphi}),I_1(\cdot,yH_\cdot^{\theta,\varphi});{\theta,\varphi},\xi)\\
&\le \inf_{({\theta,\varphi})\in\widetilde\Theta}\bar J(G^{x,{\theta,\varphi}},D^{x,{\theta,\varphi}})\ \ \mbox{({Dual})}
\end{align*}
with
$\widetilde{\Theta}:=\set{({\theta,\varphi})\in\Theta:\calX^{\theta,\varphi}(\xi)<\infty, \ \forall \xi>0}.$ Our aim is to find conditions under which we can guarantee absence of duality gap in the above formulation. In particular, if there exist an admissible pair $(\hat\pi,\kappa)$ and $(\hat\theta,\hat{\varphi}) \in \tilde\Theta$ such that $J(x;\hat\pi,\hat\kappa,D^{x,\hat{\theta,\hat\varphi}})=\bar J(G^{x,{\hat\theta,\hat\varphi}},D^{x,{\hat\theta,\hat\varphi}})$ then the strategy $(\hat\pi,\hat\kappa,D^{x,{\hat\theta,\hat\varphi}})$ is optimal. For this we consider the the linear jump-diffusion backward SDE
\begin{equation}\label{linear-BSDE}
	\begin{split}
Z_t=H_T^{\theta,\varphi} G^{\theta,\varphi}&+\int_t^T H_s^{\theta,\varphi} D_s^{\theta,\varphi}\,ds-\int_t^T\!\alpha_s\cdot\,dW_s\\
&-\int_t^T\!\bar\alpha_s\cdot\,d\bar W_s-\int_t^T\!\int_{\R^M\setminus{\set{\underline 0}}}\!\beta(s,y)\,\tilde{N}(dy,ds), \ \ t\in [0,T].
	\end{split}
\end{equation}
For the remaining part of this section we assume that for each $(\theta,\varphi)\in\Theta$ equation (\ref{linear-BSDE}) has an unique solution $(Z^{\theta,\varphi},\alpha^{\theta,\varphi},{\bar\alpha}^{\theta,\varphi},\beta^{\theta,\varphi}).$ This follows from the predictable (martingale) representation property with respect to $W,\bar W$ and $\tilde N,$ see e.g. Chapter 3 of Delong \cite{delong2013backward}. However, for CRRA preferences, existence of the solution to the above linear backward SDE can be ensured directly without using the predictable representation property. Therefore such assumption is not needed for the Examples in the next section. We have the following verification-type theorem
\begin{theorem}\label{main}
	Let Assumption \ref{depend-assumption} hold. Suppose there exist a pair $(\hat\pi,\hat\kappa)$ and $(\hat\theta,\hat{\varphi}) \in \tilde\Theta$ such that the process $Z^{\hat\theta,\hat\varphi}$ is positive and the following hold for all $t\in[0,T]$
	\begin{align}
	\sigma_t^\top\hat\pi_t&=\hat\theta^1_t+\frac{1}{Z_{t-}^{\hat\theta,\hat\varphi}}\alpha_t^{\hat\theta,\hat\varphi}, \ \ -b_t^\top\hat\kappa_t=\hat\theta_t^{2}+\frac{1}{Z_{t-}^{\hat\theta,\hat\varphi}}\bar\alpha_t^{\hat\theta,\hat\varphi},\label{optcond-pi}\\
	1-\hat\kappa_t\cdot y&=\frac{1}{\hat\varphi(t,y)}\biggl[1+\frac{\beta^{\hat\theta,\hat\varphi}(t,y)}{Z_{t-}^{\hat\theta,\hat\varphi}}\biggr]\label{opt-cond}
	\end{align}
	together with the ``complementary slackness" condition
	\begin{equation}\label{optcond-hatfpi}
	\vartheta(\zeta^{\hat{\theta}})+\hpi\cdot\zeta^{\hat\theta}=0.
	\end{equation}
	Suppose further $\kappa_{\tau_m}\cdot Y_m<1$ a.s. for $\tau_m\le T$ and $\delta_T^{\hpi,\hat\kappa,\hat D}\,ds\le x$ with
	$\hat{D}=D^{x,\hat\theta,\hat{\varphi}}.$ Then $(\hat{\pi},\hat\kappa,\hat{D})\in\mathcal A$  and this strategy is optimal.
\end{theorem}
\begin{proof}
	See Appendix.
\end{proof}

\subsection{Multiple underwriting lines with independent claims}
We will occasionally relax standing Assumption I, and suppose the following condition holds true.
\begin{Assumption}\label{indep}
	The compound Poisson processes $\sum_{\tau^j_m\le t} Y_m^j, \ j=1,\ldots,M$ are independent and each marked point process $(\tau^j_m,Y_m^j)_{m\geq 1}$ has local characteristics $(\lambda_t^j,F_t^j)$ on $[0,\infty), \ j=1,\ldots,M.$
\end{Assumption}
That is, components of the multivariate compound Poisson process are independent, so claims or jumps from any two underwriting lines can not occur simultaneously. Under this assumption, the integrals with respect to $N(dy,dt)$ satisfy
\[
\psi(t,y)\star N(dy,dt)=\sum_{j=1}^M\psi^j(t,y^j)\star N^j(dy^j,dt)
\]
where for each $j=1,\ldots,M$ we use the convention $\psi^j(t,y^j):=\psi(t,y^j\underline e^j)$ (here $\underline e^j$ denotes the unit vector with $1$ in the $j$th coordinate and $0'$s elsewhere) and $N^j(dy^j,dt)$ is the counting measure of $(\tau^j_n,Y_n^j)_{n\geq 1}$ on $(0,\infty).$   The wealth equation now reads
\begin{align*}
dV_t=V_{t-}&\biggl\{r_t\,dt+\pi_t^\top[(\mu_t-r_t\underline 1)\,dt+\sigma_t\,dW_t]+\kappa_t^\top\Bigl[p_t\,dt-b_t\,d\bar W_t\Bigr]\biggr.\\
&\biggl.-\sum_{j=1}^M\kappa_t^jy^j\star N^j(dy^j,dt)\biggr\}-D_t\,dt
\end{align*}
In this case, if no dividends are paid, then  $V_t^{\pi,\kappa,0}>0$ if $\kappa^j_{\tau_m^j}Y_m^j<1$ for $\tau_m^j\le t$ for all $j=1,\ldots,M.$ Let us denote with $\varphi(t,y)$ vectors of non-negative predictable random fields of the form $\set{\varphi^j(t,y^j)}_{1\le j\le M}.$ Then, by replacing condition (\ref{SDFcond1}) with
\begin{equation}\label{SDFcond-ind}
p_t^j+[b_t(\rho_t^\top\theta_t^1+\theta^2_t)]^j-\lambda_t^j\Exp[\varphi^j(t,Y_t^j)Y_t^j]=0, \ \ j=1,\ldots,M
\end{equation}
then the assertion of Lemma \ref{deflator} still holds true with $H^{\theta,\varphi}$ defined as
\[
dH_t=H_{t-}\Bigl\{[-r_t+\tilde\vartheta_t(\zeta_t^\theta)]\,dt-\theta_t^1\cdot dW_t-\theta_t^2\cdot\,d\bar W_t+\sum_{j=1}^M\bigl[\varphi^j(t,y^j)-1\bigr]\star\widetilde{N}^j(dy^j,dt)\Bigr\}.
\]
Here $\widetilde{N}^j(dy^j,dt):=N^j(dy^j,dt)-\lambda_t^jF_t^j(dy^j)$ for each $j=1,\ldots,M.$ For each $(\theta,\varphi)$ let $(Z^{\hat\theta,\hat\varphi},\alpha^{\hat\theta,\hat\varphi},{\bar\alpha}^{\hat\theta,\hat\varphi},\beta^{\hat\theta,\hat\varphi})$ be the solution to the  jump-diffusion backward SDE
\begin{align*}
Z_t=H_T^{\theta,\varphi} G^{\theta,\varphi}+&\int_t^T H_s^{\theta,\varphi} D_s^{\theta,\varphi}\,ds-\int_t^T\!\alpha_s\cdot\,dW_s\\
&-\int_t^T\!\bar\alpha_s\cdot\,d\bar W_s-\sum_{j=1}^M\int_t^T\!\int_{\R\setminus{\set{\underline 0}}}\!\beta^j(s,y^j)\,\tilde{N}^j(dy^j,ds), \ \ t\in [0,T].
\end{align*}
Then we have the following version of Theorem \ref{main} for the case of a multivariate compound Poisson process with independent components.
\begin{theorem}\label{main-indep}
	Under Assumption \ref{indep}, suppose there exist a pair $(\hat\pi,\hat\kappa)$ and $(\hat\theta,\hat{\varphi}) \in \tilde\Theta$ such that the process $Z^{\hat\theta,\hat\varphi}$ is positive, (\ref{optcond-pi}), (\ref{optcond-hatfpi}) and 
\begin{equation}
1-\hat\kappa_t^jy^j=\frac{1}{\hat\varphi^j(t,y^j)}\biggl[1+\frac{\beta^{\hat\theta,\hat\varphi,j}(t,y^j)}{Z_{t-}^{\hat\theta,\hat\varphi}}\biggr], \ \ j=1,\ldots,M\label{opt-cond-indep}
\end{equation}
hold a.s. for all $t\in[0,T].$ If $\kappa^j_{\tau_m^j}Y_m^j<1$ for $\tau_m^j\le T$ for all $j=1,\ldots,M$ and  $\delta^{\hpi,\hat\kappa,\hat D}_T\le x$  with
	$\hat{D}=D^{x,\hat\theta,\hat{\varphi}}.$ Then $(\hat{\pi},\hat\kappa,\hat{D})\in\mathcal A$  and this strategy is optimal.
\end{theorem}

\subsection{Precautionary earnings retention}
Here we use the definition of the dividend payout rate process $D^{x,\theta,\varphi}=I(\cdot,\Y^{\theta,\varphi}(x)H^{\theta,\varphi})$ in the dual formulation to study the impact of risk aversion, prudence, portfolio constraints and insurance risk on the the earnings retention policy of the firm. We assume $U_1$ does not depend on the time variable and $U_1=U_2\equiv U$ and the local characteristics $(\lambda_t,F_t)$ are deterministic.

Let $(\varphi,\theta)\in\Theta$ and $\xi>0$ be fixed. For simplicity, we drop dependence of $D,H$ and $\Y$ on $x,\varphi,\theta.$ Using It\^{o}'s formula and $I'(\xi)=1/U''(I(\xi)),$ $I''(\xi)=-U'''(I(\xi))/[U''(I(\xi))]^3$ we obtain
\begin{align*}
	&dI(\xi H_t)=\frac{\xi }{U''(I(\xi H_{t-}))}\,dH_t-\frac{\xi^2}{2}
	\frac{U'''(I(\xi H_{\textcolor{black}{t-}}))}{[U''(I(\xi H_{\textcolor{black}{t-}}))]^3}\,d\seq{H}_t^{\mathsf c}\\
	&+d\left\{\sum_{s\le t}\left[I(\xi H_s)-I(\xi H_{s-})-\frac{\xi }{U''(I(\xi H_{s-}))}\Delta H_{s} \right]\right\}
\end{align*}
Taking $\xi=\Y(x)=\Y^{\theta,\varphi}(x),$ and using the definition of $D=D^{x,\theta,\varphi}$ and $H=H^{\theta,\varphi}$ we get
\begin{align*}
	&dD_t=\frac{\Y(x)}{U''(D_{t-})}H_{t-}\left\{-[r_t+\vartheta_t(\zeta_t^\theta)]\,dt-\theta_t^1\cdot dW_t-\theta_t^2\cdot\,d\bar W_t+\left[\varphi(t,y)-1\right]\star\widetilde{N}(dy,dt)\right\}\\
	&-\frac{\Y(x)^2}{2}
	\frac{U'''(D_{t-})}{[U''(D_{t-})]^3}H_{t-}^2\left[|\theta^1_t|^2+|\theta^2_t|^2+2(\theta_t^1)^\top\rho_t\theta_t^2\right]\,dt\\
	&+d\biggl\{\sum_{s\le t}\Delta D_s\biggr\}-\frac{\Y(x)}{U''(D_{t-})}H_{t-}\lambda_t\Exp[\varphi(t,Y_t)-1]\,dt
\end{align*}
Now, the increments of $D$ satisfy
\[
\Delta D_s=I(\Y(x)H_{s-}\varphi(s,\Delta Y_s))-I(\Y(x)H_{s-})=I(U'(D_{s-})\varphi(s,\Delta Y_s))-D_{s-}.
\]
Since $I$ is strictly decreasing, these increments are positive (resp. negative) if $\varphi(s,y)>1$ (resp. $<1$). Rewriting jumps as integrals with respect to $N(dy,dt)$, compensating, taking expected values and rearranging, we obtain
\begin{align*}
	&\frac{d}{dt}\Exp[D_t]=\Exp\biggl[\frac{1}{\mathrm{AR}(D_t)}\left\{r_t+\vartheta_t(\zeta_t^\theta)+\lambda_t[\varphi(t,Y_{t-})-1]\right\}\biggr.\\
	&+\biggl.\frac{1}{2}
	\frac{\mathrm{AP}(D_t)}{[\mathrm{AR}(D_t)]^2}\left\{|\theta^1_t|^2+|\theta^2_t|^2+2(\theta_t^1)^\top\rho_t\theta_t^2\right\}+\lambda_t\left\{I(U'(D_{t-})\varphi(t,Y_{t-}))-D_{t-}\right\}\biggr]
\end{align*}
where $\mathrm{AR}:=-U''/U'$ and $\mathrm{AP}:=-U'''/U''$ are the absolute Arrow-Pratt coefficient of risk aversion and prudence index respectively. Since $\mathrm{AR}>0$, we see that, on average, the growth rate of $D_t^{x,\theta,\varphi}$ increases with interest rate $r_t$ and $\vartheta_t(\zeta_t^\theta).$ Moreover, if $\mathrm{AP}>0$  (resp. $<0$)  then it also responds positively (resp. negatively) to the quadratic covariation of the continuous part of state-price density $H^{\theta,\varphi}$. In presence of the insurance claims, we have in fact the following result.
\begin{theorem}
	Suppose that  $U'''>0$ (resp. $<0$) and $\varphi(t,y)>1$ (resp. $<1$) for all $y\in\supp F_t$ for all $t\in[0,T].$ Then the expected growth rate  of the dividend payout rate $D_t^{x,\theta,\varphi}$ increases with $\lambda_t.$
\end{theorem}
\begin{proof}
It suffices to prove
\[
I(U'(D)\varphi)-D+\frac{1}{\mathrm{AR}(D)}(\varphi-1)>0.
\]
for $D,\varphi\in\R_+$ fixed. Define $f(\xi):=I(U'(D)\xi).$ Suppose $U'''>0$ and $\varphi>1.$ By the mean value theorem, there exists $\varphi^*\in (1,\varphi)$ such that
\[
\frac{f(\varphi)-f(1)}{\varphi-1}=f'(\varphi^*)=I'(U'(D)\varphi^*)U'(D)=\frac{U'(D)}{U''(I(U'(D)\varphi^*))}.
\]
Since $I$ is decreasing and $U''$ is increasing, we have $I(U'(D)\varphi^*)<I(U'(D))=D$ and $U''(I(U'(D)\varphi^*))<U''(D)$ and the desired result follows. The same argument can be used if $\varphi\in (0,1)$ and $U''$ is decreasing.
\end{proof}
In particular, if $U'''>0$, that is, if the marginal utility is a convex function, and $\varphi(t,y)>1,$ as it will be the case of optimal strategies for utility functions with CRRA (see (\ref{CRRAmain}) below),  then the drift of the dividend payout process increases with the prudence index and with the  aggregate expected arrival rate of claims.

If the components of the multivariate compound Poisson process are independent, so claims or jumps from any two underwriting lines can not occur simultaneously, we have the following result.
\begin{corollary}\label{cor-FOSD}
Suppose Assumption \ref{indep} holds, $U'''>0$ (resp. $<0$), $F_t^j$ is absolutely continuous and $\varphi^j>1$ (resp. $<1$) for some $j\in\set{1,\ldots,M}.$  If $\varphi^j$ is differentiable and increasing (resp. decreasing) in $y^j$ then the expected growth rate of the dividend payout rate $D_t^{x,\theta,\varphi}$ increases (resp. decreases) with the first-order stochastic dominance of $F_t^j.$
\end{corollary}
\begin{proof}
It is a well-known fact that $F_t^j$ dominates $\tilde F_t^j$ in the sense of first-order stochastic dominance if and only if (see e.g. the book by Eeckhoudt et al \cite[Ch. 2]{eeckhoudt2011economic})
\[
\int\psi(y^j)\,F_t(dy^j)\geq \int\psi(y^j)\,\tilde F_t(dy^j)
\]
for any increasing function $\psi(y^j),$ so it suffices to prove that if $\varphi^j(t,\cdot)$ is increasing (resp. decreasing), so is
\[
\psi(y^j)=I(U'(D)\varphi^j(t,y^j))-D+\frac{1}{\mathrm{AR}(D)}(\varphi^j(t,y^j)-1)
\]
for $t\in[0,T]$ fixed. Indeed, differentiating with respect to $y^j$ we get
\begin{align*}
\psi'(y^j)&=\frac{\partial \varphi^j}{\partial y^j}\Bigl[I'(U'(D)\varphi^j(t,y^j))U'(D)+\frac{1}{\mathrm{AR}(D)}\Bigr]\\
&=\frac{\partial \varphi^j}{\partial y^j}U'(D)\Bigl[\frac{1}{U''(I(U'(D)\varphi^j(t,y^j)))}-\frac{1}{U''(D)}\Bigr].
\end{align*}
The desired result follows since $\varphi^j(t,y^j)>1,$ $I$ is decreasing and $U''$ is increasing.
\end{proof}

The intuition is that an increase in the claim frequency and/or in the first-order stochastic dominance of the claim distributions becomes a motive for precautionary earnings retention: at a given point in time, the insurer pays dividends at a lower rate compared to any time in the future. Note the prudence index enhances current earnings retention,  whereas risk aversion reduces it.

\section{Power (CRRA) utility}
In the remainder, we consider power-type utility functions with constant relative risk aversion (CRRA) of the form
\[
U_1(t,x)=U_2(x)=
\left\{
\begin{array}{ll}
\frac{x^{1-\eta}}{1-\eta}, & \ \eta\in(0,+\infty)\setminus\set{1} \\\\
\ln x, & \eta =1
\end{array}
\right.
\]
and suppose the following holds
\begin{Assumption}
	Unless $\eta=1$ (log-utility) all coefficients are non-random.
\end{Assumption}
\begin{lemma}\label{prop_CRRA}
	Suppose $(\theta,\varphi)\in {\Theta}$ are non-random. Then, we have
\[
\frac{1}{Z_{t-}^{\theta,\varphi}}\bigl(\alpha_t^{\theta,\varphi},{\bar\alpha}_t^{\theta,\varphi})^\top=\frac{1-\eta}{\eta}\theta_t, \ \ \ \frac{\beta^{\theta,\varphi}(t,y)}{Z_{t-}^{\theta,\varphi}}+1=\varphi(t,y)^{-\frac{1}{\eta}+1}
\]	
\end{lemma}
\begin{proof}
	See Appendix.
\end{proof}
Therefore, under the assumptions of the previous Lemma, conditions (\ref{optcond-pi})-(\ref{opt-cond}) become
\begin{equation}\label{CRRAmain}
\hat\theta_t^1=\eta\sigma_t^{\top}\hat\pi_t, \ \  \hat\theta_t^2=-\eta b_t^\top \hat\kappa_t, \ \ \ \hat\varphi(t,y)^{-\frac{1}{\eta}}=1-\hat\kappa_t^\top y.
\end{equation}
In what follows, for simplicity, we drop the dependence on $t\in[0,T].$ Using (\ref{CRRAmain}), we may redefine $\zeta^\theta$ in terms of $\pi$ and $\kappa$ as follows
\[
\zeta(\pi,\kappa):=r\underline 1-\mu +\eta\sigma(\sigma^\top\pi-\rho b^\top\kappa)
\]
and rewrite (\ref{SDFcond1}) and complementary slackness condition  (\ref{optcond-hatfpi}) as
\begin{equation}\label{SDFcond1-CRRA}
p+\eta b[(\sigma\rho)^\top\pi-b^\top \kappa]-\lambda\Exp\Bigl[\frac{1}{(1-\kappa\cdot Y)^\eta}Y\Bigr]=\underline 0
\end{equation}
and
\begin{equation}\label{optcond-zeta-pikappa}
\vartheta(\zeta^{\pi,\kappa})+\pi\cdot\zeta(\pi,\kappa)=0
\end{equation}
respectively. This in conjunction with Theorem \ref{main} implies the following which is our main result for CRRA preferences.
\begin{theorem}\label{main-CRRA}
	Suppose there exist a pair of processes $(\hat\pi,\hat\kappa)$ with values in $\R^d\times\R_+^M$ that solve the system of non-linear equations (\ref{SDFcond1-CRRA})- (\ref{optcond-zeta-pikappa}) with $\hat\kappa\cdot y<1$ for all $y\in\supp F.$ Suppose further $\zeta^{\hat\pi,\hat\kappa}\in\mathcal D.$ Then  $(\hat\pi,\hat\kappa)$ is optimal.
\end{theorem}
If Assumption \ref{indep} holds, that is, components of the multivariate compound Poisson process are independent, using the same argument in the proof of Lemma \ref{prop_CRRA} it can be proved easily that the optimality condition for $\hat\kappa$ and $\hat\varphi(t,y)=\set{\hat\varphi^j(t,y^j)}_{1\le j\le M}$ now becomes
\[
\hat\varphi^j(t,y^j)^{-\frac{1}{\eta}}=1-\hat\kappa_t^j y^j, \ \ j=1,\ldots M
\]
that is, $\hat\varphi^j(t,y^j)=(1-\hat\kappa_t^j y^j)^{-\eta},$ which is increasing as function of $y^j.$ Then, by Corollary \ref{cor-FOSD}, the expected growth rate of the optimal dividend payout rate for CRRA preferences increases with the first-order stochastic dominance of the claim distributions $F_t^j.$ Moreover, for this case the system of equations (\ref{SDFcond1-CRRA})  is replaced with the equations
\begin{equation}\label{SDFcond1-CRRA-indep}
p^j+\eta \left[b(\sigma\rho)^\top\pi-bb^\top \kappa\right]^j-\lambda^j\Exp\Bigl[\frac{Y^j}{(1-\kappa^jY^j)^\eta}\Bigr]=0
\end{equation}
and $\kappa^jy^j<1$ for $y^j\in\supp F^j$ for  $j=1,\ldots,M.$ We now present some examples of portfolio constraints for which solutions to (\ref{SDFcond1-CRRA}) (or (\ref{SDFcond1-CRRA-indep})) and (\ref{optcond-zeta-pikappa}) can be characterized explicitly. We first consider the unconstrained case $Q=\R^d,$ and then rectangular constraints, which include short-sale and borrowing constraints.
\subsection{Unconstrained portfolios}
The following result generalizes Theorem 4.1 of Zou and Cadenillas \cite{zou2014optimal} to the case of multiple underwriting lines with random-valued claims.
\begin{corollary}\label{main-CRRA-f0-depen}
	Suppose $Q=\R^d,$ $\sigma$ is invertible, and there exists $\hat\kappa$ such that $\hat\kappa\cdot y< 1$ on $\supp F$ satisfying the system of $M$ equations $h(\kappa)=\underline 0$ with
\begin{align*}
	h(\kappa):=p+b\Bigl[\rho^\top\sigma^{-1}(\mu-r\underline 1)-\eta (I_{M\times M}- \rho^\top\rho)b^\top\kappa \Bigr]-\lambda\Exp\Bigl[\frac{1}{(1-\kappa\cdot Y)^\eta}Y\Bigr]
\end{align*}
Then the pair $(\hat\pi,\hat\kappa)$ is optimal with
\begin{equation}\label{merton-corr}
\hat\pi=(\sigma^\top)^{-1}\left[\frac{1}{\eta}\sigma^{-1}({\mu}-r\underline 1)+\rho b^\top\hat\kappa\right].
	\end{equation}
\end{corollary}
\begin{proof}
If $Q=\R^d$ then $\vartheta=0$ and $\mathcal N=\set{\underline 0}$, so only $\zeta^{\pi,\kappa}=\underline 0$ solves (\ref{optcond-zeta-pikappa}), which is equivalent to (\ref{merton-corr}). Plugging this into (\ref{SDFcond1-CRRA}) yields the system of equations $h(\kappa)=\underline 0$, and the desired result follows.
\end{proof}
Notice the optimal portfolio equals the Merton proportion vector
\[
\pi^{\rm Merton}= \frac{1}{\eta}(\sigma\sigma^\top)^{-1}({\mu}-r\underline 1)
\]
plus the additional hedging term $ (\sigma^\top)^{-1}\rho b^\top\hat\kappa$ which helps the firm use the financial market to manage its exposure to insurance risk. We now proceed to illustrate Corollary \ref{merton-corr} numerically in the case $d=1$ and $M=2.$
\begin{example}
We assume all parameters are constant in time, and consider first an elementary example in which the the bivariate random variable $(Y_n^1,Y_n^2)$ takes values $(c_1,0), \ (0,c_2)$ and $(c_1,c_2)$ with probabilities $q_1,q_2$ and $1-(q_1+q_2)$ respectively. Then, $h^j(\kappa)$ for $j=1,2$ read
\begin{align*}
	h^1(\kappa)&=p^1+\frac{\mu-r}{\sigma}[b\rho^\top]^1-\eta \Bigl[b(I_{2\times 2}- \rho^\top\rho)b^\top\kappa \Bigr]^1\\
	&\phantom{AAA}-\lambda\left(\frac{c^1q_1}{[1-\kappa^1c^1]^\eta}+\frac{c^1[1-(q_1+q_2)]}{[1-(\kappa^1c^1+\kappa^2c^2)]^\eta}\right),\\
	h^2(\kappa)&=p^2+\frac{\mu-r}{\sigma}[b\rho^\top]^2-\eta \Bigl[b(I_{2\times 2}- \rho^\top\rho)b^\top\kappa \Bigr]^2\\
	&\phantom{AAA}-\lambda\left(\frac{c^2q_2}{[1-\kappa^2c^2]^\eta}+\frac{c^2[1-(q_1+q_2)]}{[1-(\kappa^1c^1+\kappa^2c^2)]^\eta}\right).
\end{align*}
for $\kappa\in\R_+^2$ satisfying ${c^1}\kappa^1+{c^2}\kappa^2<1.$  Figures \ref{ex1_eta4} and \ref{ex2_eta4} contain the plots of the zero-level curves $h^1(\kappa)=0$ (blue) and $h^2(\kappa)=0$ (red) for different values of $\eta$ and the following sets of parameters

\begin{itemize}
	\item[(I)] $q_1= 0.2$, $q_2=0.6$, $\mu=7\%$, $\sigma=21\%$, $r=3\%$, $\lambda=0.1$ and
\begin{equation*}
	b=\begin{bmatrix}
		0.2&0.6\\1.3&0.7
	\end{bmatrix},\quad c=\begin{bmatrix}
		3.0\\3.0
	\end{bmatrix},\quad p=\begin{bmatrix}
		0.7\\1.1
	\end{bmatrix}, \quad\rho=\begin{bmatrix}
		0.4&0.5
	\end{bmatrix}.
\end{equation*}
	\item[(II)] $q_1=0.2$, $q_2=0.7$,  $\mu=7\%$, $\sigma=21\%$, $r=3\%$ $\lambda=0.15$ and
	\begin{equation*}
		b=\begin{bmatrix}
			0.2&0.3\\0.4&0.6
		\end{bmatrix},\quad c=\begin{bmatrix}
			2.4\\1.7
		\end{bmatrix},\quad p=\begin{bmatrix}
			1.3\\0.8
		\end{bmatrix}, \quad\rho=\begin{bmatrix}
			-0.2&0.3
		\end{bmatrix}.
	\end{equation*}
\end{itemize}
\begin{figure}[t!]
	\centering
	\includegraphics[scale=0.34]{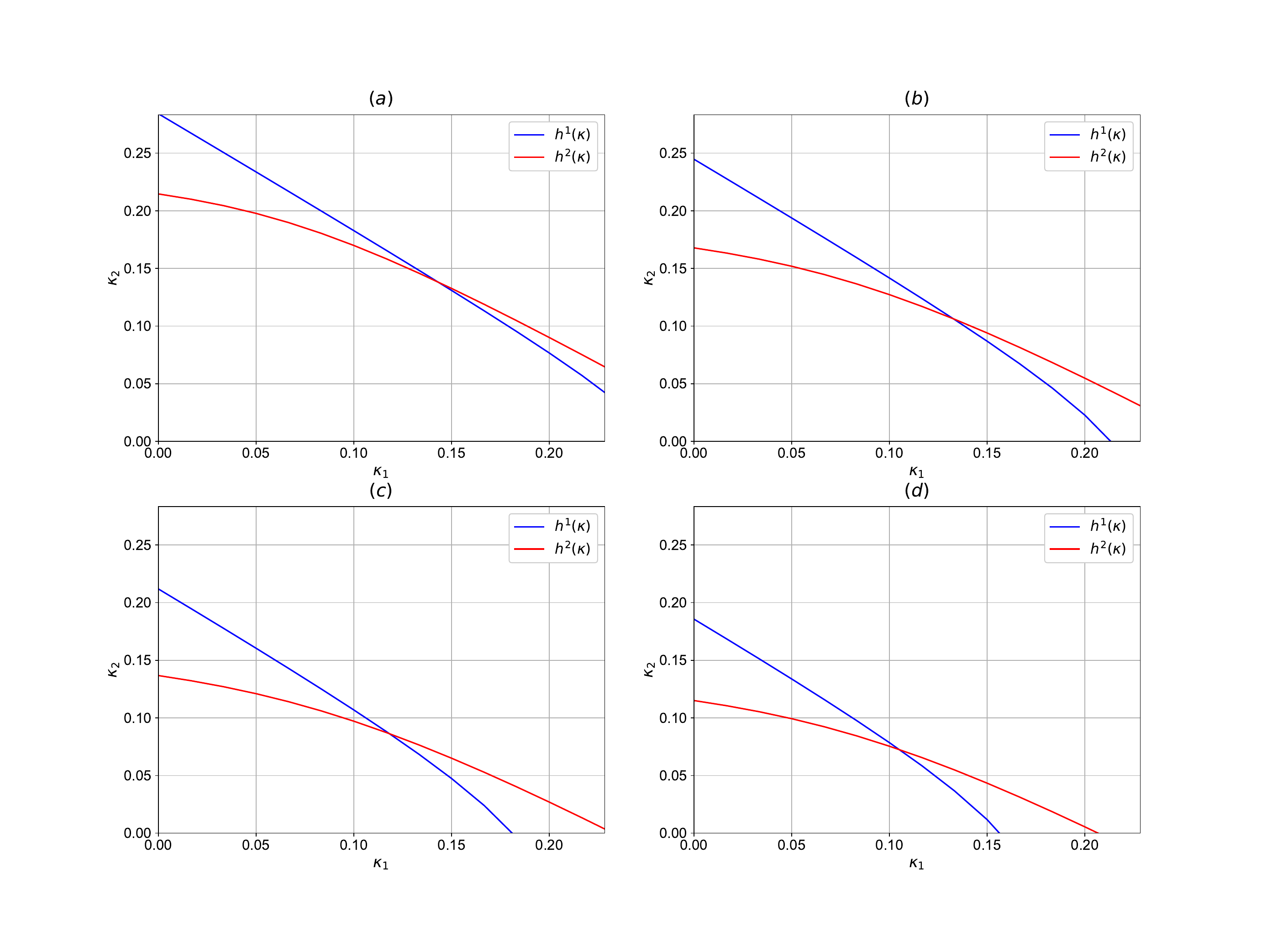}
	\caption{Level curves $h^j(\kappa)=0$ for $j=1,2$ and parameter set (I). The values of $\eta$ are (a) 0.7, (b) 1.2, (c) 1.7 and (d) 2.2.}\label{ex1_eta4}
\end{figure}
\begin{figure}[h!]
	\centering
		\includegraphics[scale=0.34]{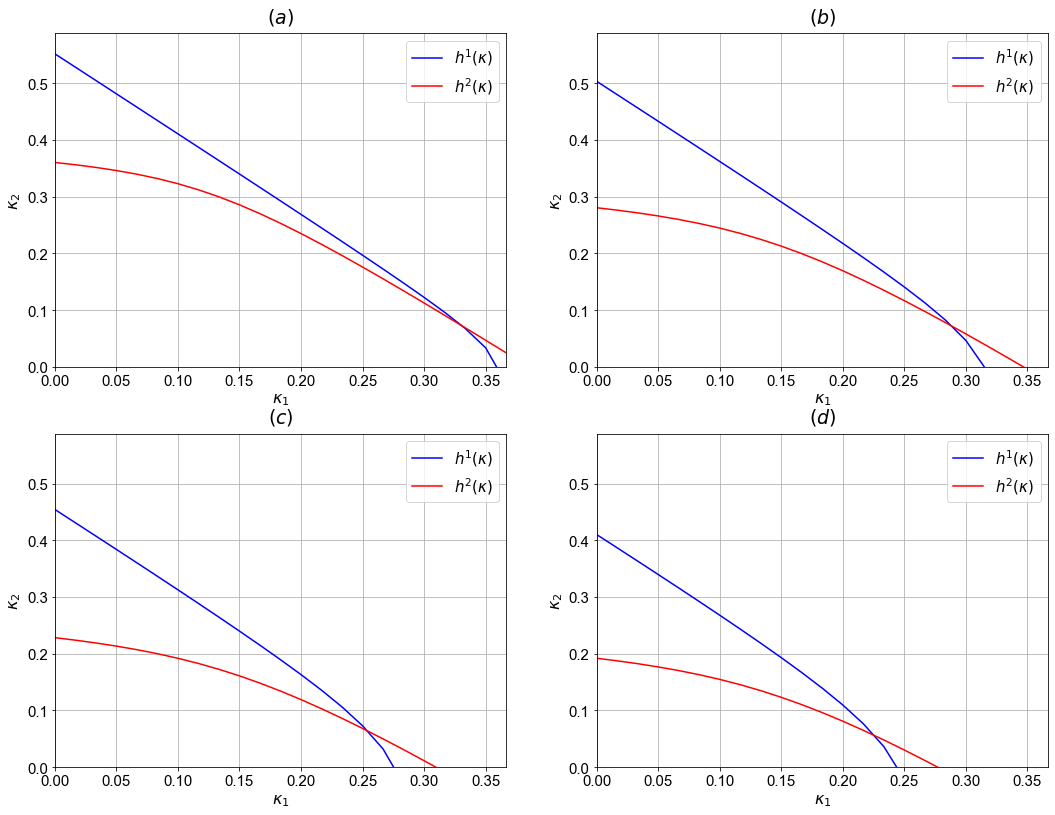}
	\caption{Level curves $h^j(\kappa)=0$ for $j=1,2$ and parameter set (II). The values of $\eta$ are (a) 1.2, (b) 1.7, (c) 2.2 and (d) 2.7.}\label{ex2_eta4}
\end{figure}
Figure \ref{dep_eta} contain the plots of optimal $\hat\kappa$ as a function of $\eta\in[0.3,4].$ We see that both $\hat\kappa^1$ and $\hat\kappa^2$ decrease to zero, and the respective solvency thresholds increase, for high values of risk aversion coefficient $\eta.$ However, this behavior differs for low risk aversion levels due to the different signs of correlation coefficients: for parameter set (I) $\hat\kappa^1$ increases and $\hat\kappa^2$ decreases, while for parameter set (II) $\hat\kappa^1$ decreases and $\hat\kappa^2$ increases.
\begin{figure}[ht!]
	\centering
	\subfloat{\includegraphics[scale=0.28]{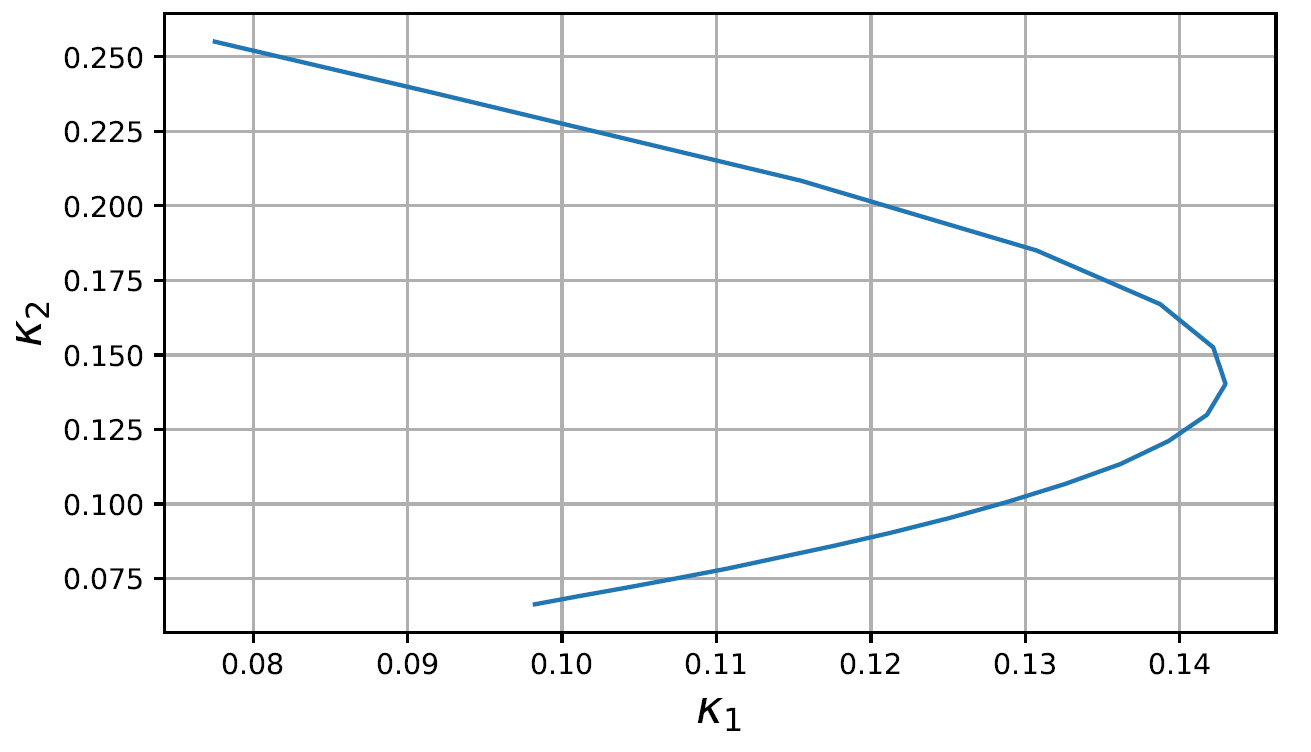}}%
	\qquad
	\subfloat{\includegraphics[scale=0.3]{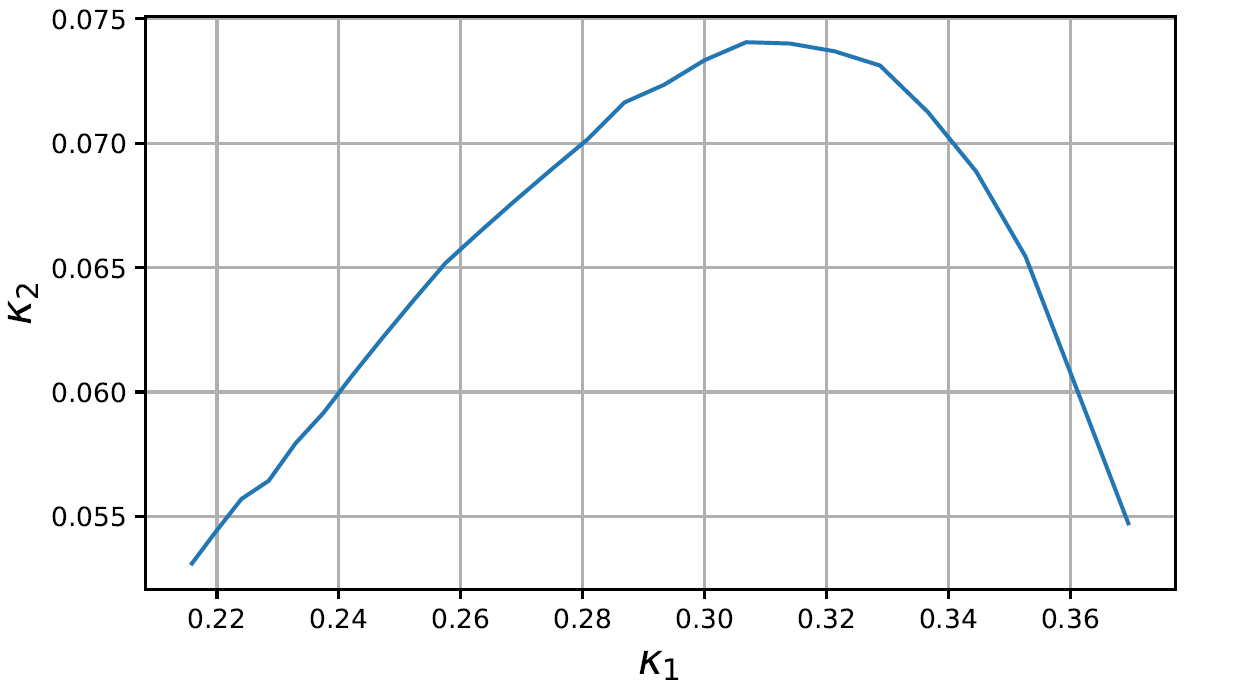}}%
	\caption{Optimal $\hat\kappa$ as a function of $\eta$ for parameter sets (I) and (II).}%
	\label{dep_eta}%
\end{figure}

\end{example}
For the next example we assume a more general setting for the multivariate compound Poisson process by employing the dependence characterization of its components via L\'{e}vy copulas and tail integrals, e.g. see Kallsen and Tankov \cite{kallsen-tankov}. Indeed, Sklar's Theorem for multivariate L\'{e}vy processes ensures that there exists a L\'{e}vy copula $\mathfrak C:[0,+\infty]^M\to [0,+\infty]$  such that the L\'{e}vy measure $\nu(dy)$ of the multivariate compound Poisson process $\sum_{\tau_n\le t}Y_n$ satisfies
\begin{equation}\label{tail-CPP}
\nu\left([y^1,+\infty)\times\cdots\times [y^M,+\infty)\right)=\mathfrak{C}\left(\lambda F^1 ([y^1,+\infty)),\ldots,\lambda F^M ([y^M,+\infty)\right)
\end{equation}
for $y^j>0,$ where $F^j$ denotes the marginal distributions of the components $Y_n^j.$ We assume the $[0,\infty)^M$-valued random variables $Y_n$ do not have atoms in $\underline 0$ i.e. $\Prob(Y_1=\underline 0)=0.$ However, as mentioned above, it is possible that single claims in one of the lines occur, in which case the distributions of the severities $Y_n^j$ may not be absolutely continuous and have positive masses at $0,$ that is, $F_{Y^j}(0)=\Prob(Y_n^j=0)$ may not be zero. Again, we focus on the case $M=2,$ and assume the maximum loss condition
\begin{equation}\label{policy-limit}
	\supp F\subseteq [0,c^1]\times[0,c^2]
\end{equation}
holds for some positive numbers $c_1,c_2.$ To ensure this, for simplicity, we assume the marginal severities satisfy the policy limit condition
\[
\Prob(Y_n^j\le y^j|Y_n^j>0)=\Prob(Z_m^j\wedge c^j\le y^j)
\]
where $Z^j$ is absolutely continuous with density $f^j,\ j=1,2.$ Then, the joint density of $\nu(dy)$ is given as
\begin{align*}
&f(y^1,y^2)\\
&=\left\{
\begin{array}{ll}
\lambda^1\lambda^2\frac{\partial^2\mathfrak C}{\partial y^1\partial y^2}\left(\lambda^1\bar  F_{Z^1} (y^1),\lambda^2\bar  F_{Z^2}(y^2)\right)f^{1}(y^1)f^2(y^2), \  &y^1<c^1, \ y^2<c^2\\\\
\lambda^1\lambda^2\frac{\partial^2\mathfrak C}{\partial y^1\partial y^2}\left(\lambda^1\bar  F_{Z^1} (y^1),\lambda^2\bar  F_{Z^2}(c^2)\right)f^{1}(y^1)\bar  F_{Z^2} (c^2), \  &y^1<c^1, \ y^2=c^2\\\\
\lambda^1\lambda^2\frac{\partial^2\mathfrak C}{\partial y^1\partial y^2}\left(\lambda^1\bar  F_{Z^1} (c^1),\lambda^2\bar  F_{Z^2}(y^2)\right)\bar  F_{Z^1} (c^1)f^2(y^2), \  &y^1=c^1, \ y^2<c^2\\\\
\lambda^1\lambda^2\frac{\partial^2\mathfrak C}{\partial y^1\partial y^2}\left(\lambda^1\bar  F_{Z^1} (c^1),\lambda^2\bar  F_{Z^2}(c^2)\right)\bar  F_{Z^1} (c^1)\bar  F_{Z^2} (c^2), \  &y^1=c^1, \ y^2=c^2\\
\end{array}
\right.
\end{align*}
where $\bar F_{Z^j}$ denotes survival function. Again, we restrict $\kappa$ so that ${c^1}\kappa^1+{c^2}\kappa^2<1.$ Note this implies that the wealth process must be larger than $L^1c^1+L^2c^2$ which is obviously quite restrictive from the practical point of view. Later we relax this condition by considering multivariate compound Poisson process with independent components.
\begin{example}\label{clayton}
We assume a Clayton copula of the form
\[
\mathfrak C(u,v)=(u^{-\delta}+v^{-\delta})^{-\frac{1}{\delta}}, \ u,v>0
\]
with dependence parameter $\delta>0.$ Then
		\[
		\frac{\partial^2\mathfrak C}{\partial u\partial v}(u,v)=(\delta+1)(uv)^\delta(u^\delta+v^\delta)^{-\frac{1}{\delta}-2}.
		\]
We also assume 	$Z_n^1\sim\textrm{Exp}(\theta)$ and $Z_n^2\sim\textrm{Weibull}(\varsigma,\varrho)$  with density functions
		\begin{equation}\label{exp_dist}
		f^1(z)=\frac{1}{\theta}e^{-z/\theta},\quad\text{for }z\geq0,
		\end{equation}
		and
		\begin{equation}\label{Weibull}
		f^2(z)=\left(\frac{\varsigma}{\varrho}\right)\left(\frac{z}{\varrho}\right)^{\varsigma-1}\exp\left\{-\left(\frac{z}{\varrho}\right)^\varsigma\right\},\quad \text{for }z\geq0.
		\end{equation}
		Again, we restrict to the case of one risky asset. Figure \ref{figdep} shows the plots of the zero-level curves for the following specifications: 	
		$\mu=7\%$, $\sigma=21\%$, $r=3\%$, $\theta=2$, $\varrho=2$, $\varsigma=0.5$ and
		\begin{equation*}
			b=\begin{bmatrix}
				0.2&0.6\\1.3&0.7
			\end{bmatrix},\quad c=\begin{bmatrix}
				3.0\\3.0
			\end{bmatrix},\quad p=\begin{bmatrix}
				0.5\\0.4
			\end{bmatrix}, \quad\rho=\begin{bmatrix}
				0.3&0.5
			\end{bmatrix}\quad\lambda=\begin{bmatrix}
				1.1\\0.1
			\end{bmatrix}.
		\end{equation*}
Table \ref{Tabla_delta_eta} contains the optimal strategies $(\hat\pi,\hat\kappa).$ In this case, we see that as the dependence parameter $\delta$ increases, since correlations $\rho^1,\rho^2$ are positive, insurance exposure can be hedged away partially more efficiently, so both $\hat\kappa^1$ and $\hat\kappa^2$ increase, and so does $\hat\pi.$ However, as risk aversion $\eta$ increases, both $\hat\kappa^1$ and $\hat\kappa^2$ decrease, for a fixed level of dependence $\delta.$    
\begin{figure}[ht!]%
	\centering
	\subfloat[$\delta=1.3, \ \eta=1.2$]{\includegraphics[scale=0.27]{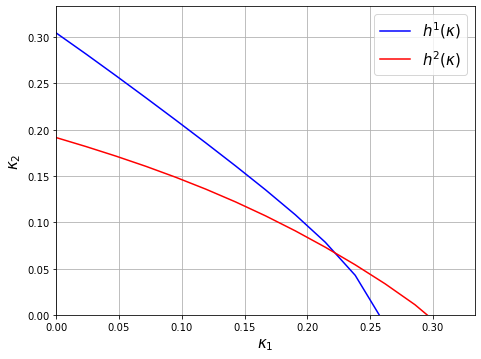}}%
	\ 
	\subfloat[$\delta=1.3, \ \eta=1.5$]{\includegraphics[scale=0.27]{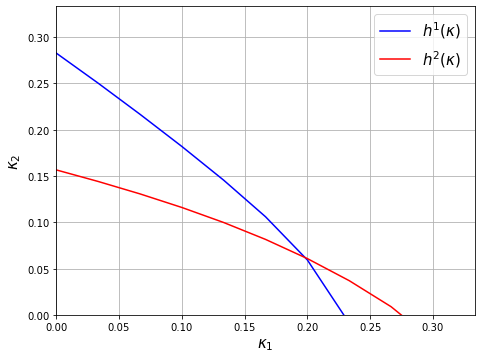}}%
	\ 
	\subfloat[$\delta=1.3, \ \eta=1.8$]{\includegraphics[scale=0.27]{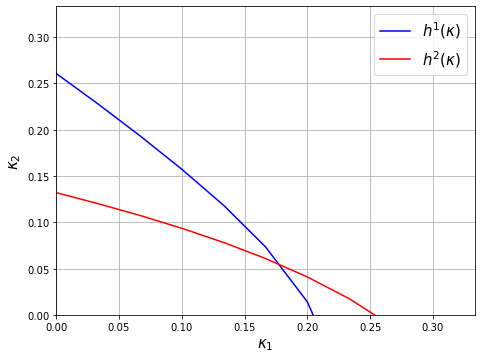}}\\
	\subfloat[$\delta=1.8, \ \eta=1.2$]{\includegraphics[scale=0.27]{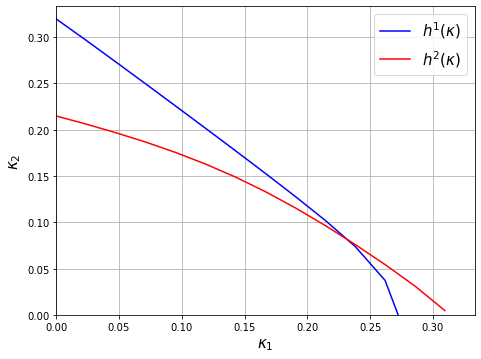}}%
	\ 
	\subfloat[$\delta=1.8, \ \eta=1.5$]{\includegraphics[scale=0.27]{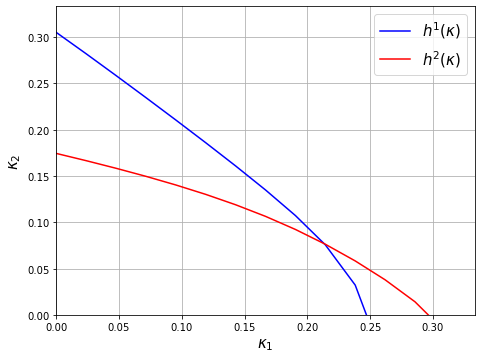}}%
	\ 
	\subfloat[$\delta=1.8, \ \eta=1.8$]{\includegraphics[scale=0.27]{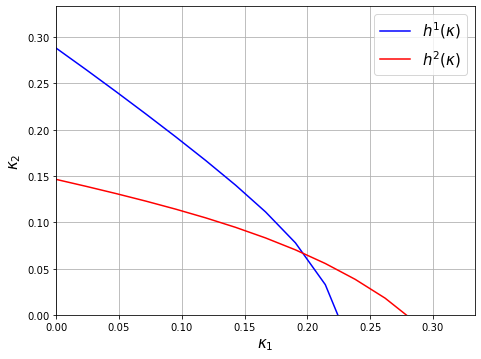}}\\
	\subfloat[$\delta=2.3, \ \eta=1.2$]{\includegraphics[scale=0.27]{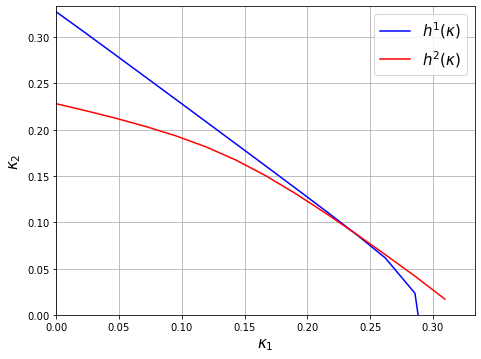}}%
	\ 
	\subfloat[$\delta=2.3, \ \eta=1.5$]{\includegraphics[scale=0.27]{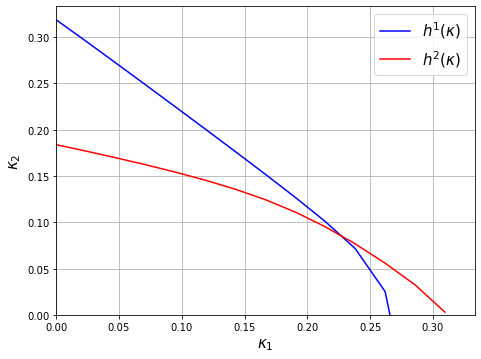}}%
	\ 
	\subfloat[$\delta=2.3, \ \eta=1.8$]{\includegraphics[scale=0.27]{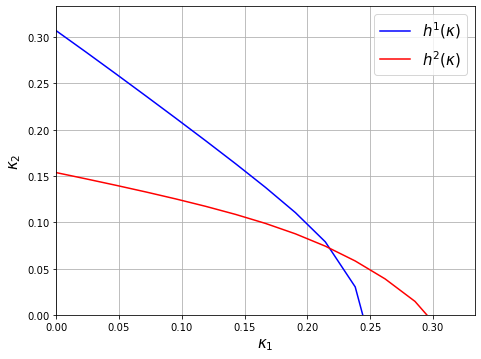}}
	
	\caption{Level curves $h^j(\kappa)=0$ for $j=1,2$ and different values of $\delta$ and $\eta,$ for the bivariate model with $Z^1\sim \textrm{Exp}(2)$, $Z^2\sim \textrm{Weibull}(0.5,2)$.}%
	\label{figdep}%
\end{figure}

\begin{table}[ht!]
	\centering
	\begin{tabular}{|c|c|c|c|c|}
		\hline
		$\eta$&$\delta$&$\hat\kappa^1$&$\hat\kappa^2$&$\hat\pi$\\
		\hline
		\multirow{3}{*}{1.2}&1.3&0.2227&0.0668&1.3731\\
		\cline{2-5}
		&1.8&0.2297&0.0840&1.4457\\
		\cline{2-5}
		&2.3&0.2356&0.0907&1.4797\\
		\hline
		\multirow{3}{*}{1.5}&1.3&0.1980&0.0625&1.1647\\
		\cline{2-5}
		&1.8&0.2133&0.0772&1.2424\\
		\cline{2-5}
		&2.3&0.2277&0.0852&1.2953\\
		\hline
		\multirow{3}{*}{1.8}&1.3&0.1788&0.0543&1.0021\\
		\cline{2-5}
		&1.8&0.1972&0.0668&1.0773\\
		\cline{2-5}
		&2.3&0.2189&0.0716&1.1315\\
		\hline
	\end{tabular}
	\caption{Optimal strategies for the bivariate model with $Z^1\sim \textrm{Exp}(2)$, $Z^2\sim \textrm{Weibull}(0.5,2)$.}
	\label{Tabla_delta_eta}
\end{table}	
\end{example}

For the case of business lines with independent compound Poisson processes, we have the following result. The proof is the same as in Corollary \ref{main-CRRA-f0-depen}.
\begin{corollary}
	Suppose $Q=\R^d$ and that Assumption \ref{indep} also holds. If there exists $\hat\kappa$ satisfying $\hat\kappa^jy^j\le 1$ for $y^j\in\supp F^j$ and the system of $M$ equations $h(\kappa)=\underline 0$ with
	\begin{equation}\label{eq-kappa-ind}
	h^j(\kappa):=\,p^j+\eta  \left[b\Bigl(\rho^\top\sigma^{-1}\Bigl[\frac{1}{\eta}({\mu}-r\underline 1)+\sigma\rho b^\top{\kappa}\Bigr]-b^\top\kappa\Bigr)\right]^j-\lambda^j\Exp\Bigl[\frac{Y^j}{(1-{\kappa^j}Y^j)^\eta}\Bigr]
	\end{equation}
	for $j=1,\ldots,M,$ then $(\hat\pi,\hat\kappa)$ is optimal, with $\hat\pi$ given by (\ref{merton-corr}).
\end{corollary}
For the numerical examples, we can relax the constraint on the insurance control variable $\kappa.$ 
Namely, we restrict $\kappa$ to the hyper-rectangle $\prod_{j=1}^M\bigl[0,\frac{1}{c^j}\bigr],$ which weakens the no-bankruptcy constraint significantly. Indeed, the wealth process must be larger than $L^jc^j$ for all $j=1,\ldots,M,$ that is, the value of the total reserve is larger than the maximum loss in each of the underwriting lines. This is much more reasonable for non-life multiline insurers, yet correlations among the diffusion parts of the insurance risk process allow us to model interdependence between variations of claims paid and premiums received, see also Remark \ref{DepVsIndep-claims} above.

\begin{example}\label{ex_indep}
To illustrate this result, we suppose again {$M=2$ and $d=1,$$Y_n^j=Z_n^j\wedge c^j,$ $j=1,2$ and $Z_n^1\sim \textrm{Exp}(2.5)$ and $Z_n^2\sim \textrm{Weibull}(1.1,0.7)$. Figure \ref{fig1} (a) contains the plots of the zero-level curves $h^1(\kappa)=0$ (blue) and $h^2(\kappa)=0$ (red) for the following parameters: $\eta=1.7$, $\mu=5\%$, $\sigma=21\%$, $r=3\%$,
	\[
	b=\begin{bmatrix}1.0&0.5\\1.4&0.7\end{bmatrix}, \ \lambda=\begin{bmatrix} 0.05\\0.10\end{bmatrix}, \ c=\begin{bmatrix}3.0\\3.0\end{bmatrix}, \ p=\begin{bmatrix} 0.7\\1.0\end{bmatrix} \ \text{ and } \ \rho=\begin{bmatrix}0.4&0.5\end{bmatrix}.
	\]
	The parameters of (b) are the same of (a) but with lower risk-aversion parameter $\eta=1.10$. The parameters of (c) are the same of (a) but with $\lambda^1=0.01$. The parameters of (d) are the same of (a) but with $\lambda^2=0.01$. 	Table \ref{tab:my_label_M2}} reports the optimal values of $\kappa$ and the porfolio proportion $\pi$ for these and other values of $\eta,$ $\lambda^1$ and $\lambda^2.$ We see that if either $\lambda^1$ or $\lambda^2$ increases, the corresponding optimal liability ratio decreases, while the other one increases. The intuition is that if correlations with financial market are positive for both lines,  an increase in the claim frequency of an underwriting line moves its optimal solvency threshold in the same direction, while the optimal solvency threshold for the other line decreases.
\begin{center}
	\begin{figure}[ht!]
		\includegraphics[width=\textwidth]{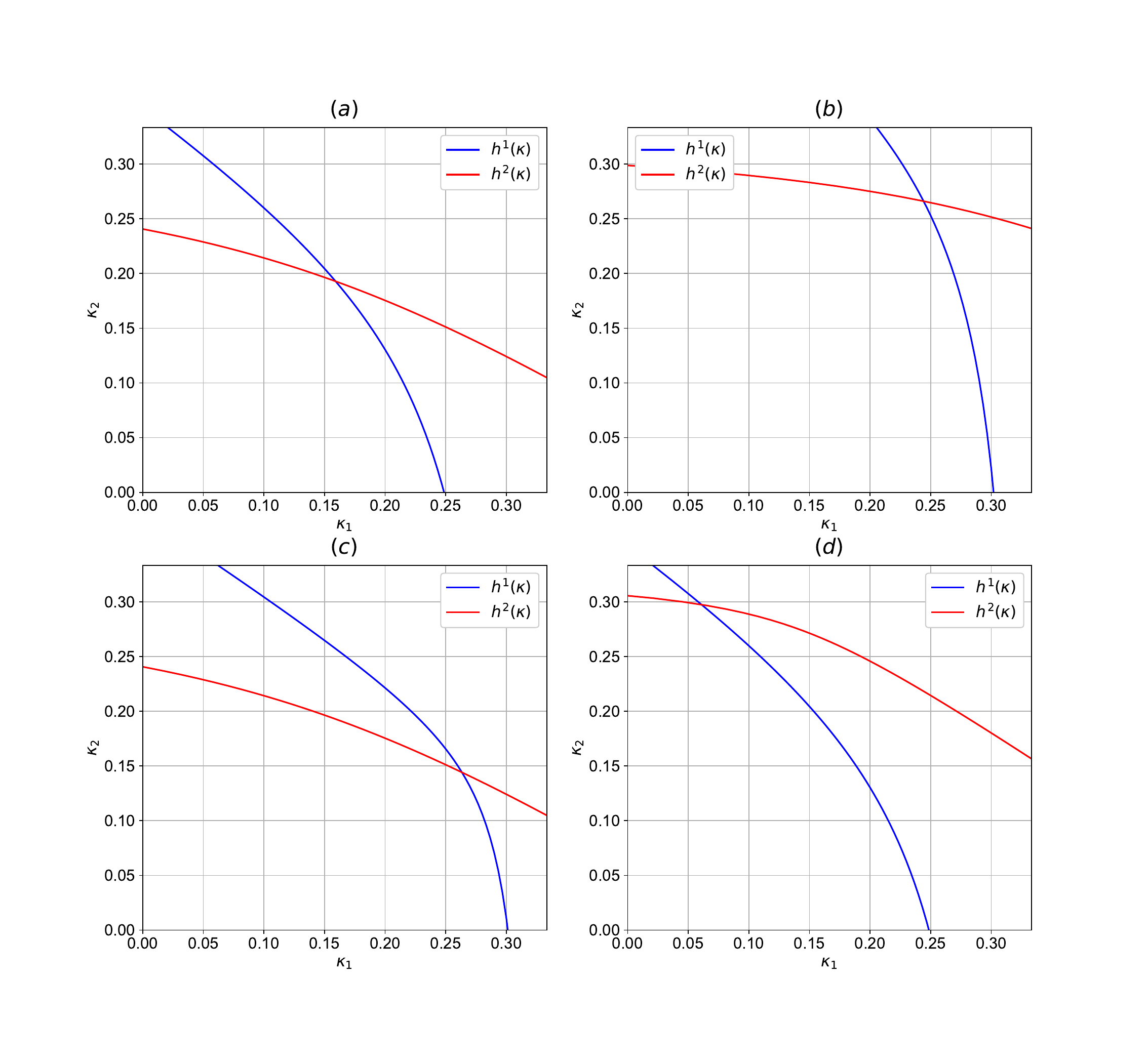}
		\caption{Level curves $h^j(\kappa)=0$  for $j=1,2$ and independent compound Poisson processes with $Z_n^1\sim \textrm{Exp}(2.5)$ and $Z_n^2\sim \textrm{Weibull}(1.1,0.7)$.}
		\label{fig1}
	\end{figure}	
\end{center}

\begin{table}[th!]
	\centering
	\begin{tabular}{|c|c|c|c|c|c|}
		\hline
		$\eta$&$\lambda^1$&$\lambda^2$& $\hat{\kappa}^1$&$\hat{\kappa}^2$ &$\hat{\pi}$ \\
		\hline
		1.1&0.05&0.10&0.2207&0.2593&2.2193\\
		1.6&0.05&0.10&0.1531&0.1964&1.6087\\
		2.1&0.05&0.10&0.1159&0.1557&1.2500\\
		2.6&0.05&0.10&0.1139&0.1537&1.1936\\
		3.0&0.05&0.10&0.0795&0.1134&0.8892\\
		1.7&0.05&0.10&0.1439&0.1868&1.5221\\
		1.7&0.03&0.10&0.1868&0.1695&1.5805\\
		1.7&0.01&0.10&0.2519&0.1405&1.6558\\
		1.7&0.05&0.05&0.1063&0.2294&1.5905\\
		1.7&0.05&0.03&0.0839&0.2549&1.6316\\
		1.7&0.05&0.01&0.0527&0.2929&1.6994\\
		\hline
	\end{tabular}
	\caption{Optimal $\hat{\kappa}$ and $\hat{\pi}$ for independent compound Poisson processes, for different values of $\eta,\lambda^1$ and $\lambda^2.$}
	\label{tab:my_label_M2}
\end{table}
\end{example}

\subsection{Rectangular constraints}
Suppose now that $Q=\prod_{k=1}^{K}I_k$ with $I_k=[\underline q^k,{\overline q}^k],$ $-\infty\le\underline q^k\le 0\le\overline q^k\le\infty$ and with the understanding that $I_k$ is open on the right (resp. left) if $\overline q^k=\infty$ (resp. $\underline q^k=-\infty$). Then $\mathcal N=\R^K$ and
\[
\vartheta(\zeta)=\sum_{k=1}^d\overline q^k(\zeta^k)^-\underline q^k(\zeta^k)^+
\]
if all the $\underline q_k$ and ${\overline q}_k$ are finite. More generally,
\[
\mathcal N=\set{\zeta\in\R^K:\zeta^i\geq 0 \mbox{ if } \overline q^i=\infty, \ \zeta^k\le 0 \mbox{ if } \underline q^k=-\infty, \mbox{ for some }i,k=1,\ldots,K}
\]
and the formula for $\vartheta(\zeta)$ remains valid. For the sake of illustration, we consider the case of one-risky asset with both short-selling and borrowing constraints, and the case with several risky assets with prohibition of short-selling $Q=[0,\infty)^K$.

\begin{example}[Short-selling and borrowing constraints, one risky asset] We asumme $K=1$ and $Q=[\underline q,\overline q]$ with $-\infty<\underline q\le 0\le \overline q<\infty.$ For this case, complementary slackness condition (\ref{optcond-zeta-pikappa}) becomes
\[
\overline q\zeta(\pi,\kappa)^--\underline q\zeta(\pi,\kappa)^+=0.
\]
The following three possible solutions can be singled out.
\begin{enumerate}
	\item Let $(\hpi,\hat\kappa)$ be as in Corollary \ref{main-CRRA-f0-depen}. If $\hpi\in[\underline q,\overline q]$ then $(\hpi,\hat\kappa)$ is optimal.
	
	\item Let $\hat\kappa$ be such that $\hat\kappa\cdot y< 1$ for all $y\in\supp F$ and solves the system of $M$ equations $g(\hat\kappa)=p+\overline q \eta \sigma b\rho^\top$ (resp. $p+\underline q \eta \sigma b\rho^\top$) with
\[
g(\kappa):=	\eta bb^\top \kappa+\lambda\Exp\Bigl[\frac{1}{(1-\kappa\cdot Y)^\eta}Y\Bigr].
\]
If $\eta\sigma\rho b^\top\hat\kappa+\mu-r>\eta\sigma^2\overline q$ (resp. $<\eta\sigma^2\underline q$) then $(\overline q,\hat\kappa)$ (resp. $(\underline q,\hat\kappa)$) is optimal.
\end{enumerate}
\end{example}

\begin{example}[Prohibition of short-selling, multiple risky assets]	
Finally, we consider the case $Q=[0,\infty)^K.$ In this case, we have $\N=Q$ and $\vartheta_t\equiv 0$ on $\N.$ Complementary slackness condition (\ref{optcond-zeta-pikappa}) simplifies into $\pi^\top\zeta(\pi,\kappa)=0.$
We can single out the following cases
\begin{enumerate}
  \item Let $(\hpi,\hat\kappa)$ be as in Corollary \ref{main-CRRA-f0-depen}. If $\hpi\in\R_+^K$ then $(\hpi,\hat\kappa)$ is optimal.
  \item Suppose there exists $\hat\kappa$ satisfying $\hat\kappa\cdot y< 1$ for all $y\in\supp F$ and solution to the system of $M$ equations $g(\hat\kappa)=p.$

If $r\underline 1-\mu -\eta\sigma\rho b^\top\hat\kappa\in\R_+^d$ then the pair $(\underline 0,\hat\kappa)$ is optimal. In this case, the expected returns of the risky assets are too low, so it is optimal to invest all the underwriting profits in the risk-free asset.

  \item Let $\underline e^i$ denote the unit vector in the $i$-th coordinate. Suppose that for some $i\in\set{1,\ldots,d}$ there exists $(\hat\beta,\hat\kappa)\in\R_+^{1+M}$ satisfying $\hat\kappa\cdot y< 1$ for all $y\in\supp F$ and solution to the system of $1+M$ equations\footnote{Here, $[\cdot]^i$ denotes the $i$-th column.}
  \begin{align*}
  g(\kappa)-\beta\eta [b\rho^\top\sigma^\top]^i&=p\\
  \beta\eta[\sigma\sigma^\top]^{ii}-\eta[\sigma\rho b^\top\kappa]^i&=\mu^i-r.
  \end{align*}
  If $\beta\eta[\sigma\sigma^\top]^{ji}-\eta[\sigma\rho b^\top\kappa]^j>\mu^j-r$ for all $j\neq i,$ then $(\hat\beta\underline e^i,\hat\kappa)$ is optimal. That is, it is optimal to invest the fraction $\hat\beta$ of the underwriting profits in the risky asset $S^i,$ and the fraction $1-\hat\beta$ in the risk-free asset. If $\hat\beta>1$ the position in the risky asset must be financed by borrowing at the risk-free rate $r.$
  
\end{enumerate}
\end{example}



\section{Conclusions}
Insurance companies are expected to be exposed to the financial sector since they invest the
proceeds of the policyholders’ premiums in the financial market. The growing expansion of financial companies that conduct insurance business into investment-bank-like activities, especially through financial conglomerates, has considerably deepened the exposure of the insurance industry to financial risks. However, this has also created complex incentive problems when different parts of a conglomerate pursuing activities with different risk profiles use the same capital base. This clearly underlines the importance of properly understanding the financial risks faced by insurance firms, especially those with investment activities, and of considering the various interrelations between financial assets and underwriting risks.

In this paper we have extended the classical Lagrangian convex duality approach to solve the portfolio allocation problem of a multiline insurance firm. The particular structure of cointegration between investments and insurance liabilities enables us to solve fully characterize optimal ALM strategies for CRRA power preferences. In particular, we prove that both financial and multivariate underwriting risks can be hedged away partially in an efficient manner in the face of extreme events and frictions. This result allows to address important practical issues such as the sensitivity of optimal policies with respect to the risk aversion and model parameters.

The case in which the multivariate compound Poisson process with independent components that never jump together is of particular importance since the solvency constraint can be significantly weakened, yet correlations among the diffusion parts of the insurance risk process still allow to model interdependence between variations of claims paid and premiums received. This sheds light on the relevance of our findings on a non-technical level. Our numerical examples also show the impact of cointegration on investment-insurance ALM with multiple (dependent and independent) sources of insurance risk.

\appendix

\section{Proofs}
	
\begin{proof}[Proof of Lemma \ref{deflator}]
Denote $V:=V^{\pi,\kappa,D}$ and $H:=H^{\theta,\varphi}.$ Using integration-by-parts formula for jump-diffusions we get
	\begin{equation*}
	d(V_tH_t)=H_{t-}dV_t+V_tdH_{t-}+d\langle V^c,H^c \rangle+d\Bigl[\sum_{s\leq t}\Delta H_s \Delta V_s \Bigr].
	\end{equation*}
	Here $\langle V^c,H^c\rangle$ denotes the quadratic co-variation process of the continuous parts of $H$ and $V$. Then
	\begin{align*}
	\frac{d(V_tH_t)}{H_{t-}V_{t-}}&=[r_t+\pi_t\cdot(\mu_t-r_t\underline 1)+\kappa_t\cdot p_t]\,dt+\pi_t^\top\sigma_t\,dW_t\biggr.\\
	&\phantom{AA}\biggl.-\kappa_t^\top\Bigl[b_t\,d\bar W_t+y\star N(dy,dt)\Bigr]-[r_t+\vartheta_t(\zeta_t^\theta)]\,dt\\
	&\phantom{AA}\biggl.-\theta_t^1\cdot dW_t-\theta_t^2\cdot\,d\bar W_t+\left[\varphi(t,y)-1\right]\star\widetilde{N}(dy,dt)-\left[\pi_t^\top\sigma_t\theta^1-\kappa_t^\top b_t\theta^2\right]dt\\
	&\phantom{AA}\biggl.-\kappa_t^\top y[\varphi(t,y)-1]\star N(dy,dt)-\left[\pi_t^\top\sigma_t\rho_t\theta_t^2-\kappa_t^\top b_t\rho_t^\top\theta_t^1\right]dt -\frac{D_t}{V_{t-}}dt.
	\end{align*}
Compensating the integrals with respect the jump measure $N(dy,dt)$ and using condition \eqref{SDFcond1} we get
	\begin{align*}
	\frac{d(V_tH_t)}{H_{t-}V_{t-}}&=\bigl[-\pi_t\cdot\zeta_t^\theta-\vartheta_t(\zeta_t^\theta)-\frac{D_t}{V_{t-}}\bigr]dt+[\pi_t^\top\sigma_t-\theta_t^1]\,dW_t\\
	&\phantom{AA}\biggl.-[\kappa_t^\top b_t+\theta_t^2]d\bar{W}_t+[\varphi(t,y)(1-\kappa_t^\top y)-1]\star\widetilde{N}(dy,dt).
	\end{align*}
By definition of $\vartheta_t$ we have $-\pi_t\cdot\zeta_t^\theta-\vartheta_t(\zeta_t^\theta)\le 0$ for all $t\in [0,T].$ Integrating we obtain that  $H_t^{\theta,\varphi}V_t^{\pi,\kappa,D}+H_t^{\theta,\varphi}D_t$ is a non-negative local-martingale. In particular, by Fatou's lemma it is a super-martingale, and the desired result follows.
\end{proof}

\begin{proof}[Proof of Theorem \ref{main}]
The proof adapts the arguments of Michelbrik and Le \cite[Theorem 1]{michelbrink} to the setting with insurance risk and portfolio constraints. Using It\^o's formula for jump-diffusion processes with the function $1/x$ and the process $H:=H^{\hat\theta,\hat\varphi}$ we get
	\begin{align*}
	d\left(\frac{1}{H_t}\right)&=-\frac{1}{H^2_{t-}}dH_t+\frac{1}{H^3_{t-}}d\langle H^c\rangle_t+\left[\frac{1}{\varphi(t,y)}-1+\hat\varphi(t,y)-1\right]\star N(dy,dt)\\
	&=\frac{1}{H_{t-}}\left\{ \left[r_t+\tilde{\vartheta}_t(\zeta^{\hat\theta})+|\hat\theta^1_t|^2+|\hat\theta^2_t|^2+2(\hat\theta_t^1)^\top\rho_t\hat\theta_t^2\right]dt+\hat\theta^1\cdot dW_t\right.\\
	&\phantom{AAAAAA}\biggl.+\hat\theta^2\cdot d\bar{W}_t+\Bigl[\frac{1}{\hat\varphi(t,y)}-1 \Bigr]\star N(dy,dt)+\lambda_t\Exp[\hat\varphi(t,Y_t)-1]\,dt\biggr\}.
	\end{align*}
Recall that $Y_0:=0$ and $Y_t:=Y_n$ if $t\in(\tau_{n-1},\tau_n].$ For simplicity, we use the notation $\alpha=\alpha^{\hat\theta,\hat\varphi},\bar\alpha=\bar\alpha^{\hat\theta,\hat\varphi}, \beta=\beta^{\hat\theta,\hat\varphi}$ and $Z:=Z^{\hat\theta,\hat\varphi}.$ Using integration-by-parts formula for jump-diffusion processes, the differential of the process $Z/H$ satisfies
	\begin{align*}
&d\Bigl(\frac{Z_t}{H_t}\Bigr)=Z_{t-}d\Bigl(\frac{1}{H_{t-}}\Bigr)+\frac{1}{H_{t-}}dZ_t+d\Bigl< Z^c,\frac{1}{H^c}\Bigr>_t+\frac{1}{H_{t-}}\beta(t,y)\left[ \frac{1}{\hat\varphi(t,y)}-1\right]\star N(dy,dt)\\
	&=\frac{Z_{t-}}{H_{t-}}\left\{\left[r_t+\tilde{\vartheta}_t(\zeta^{\hat\theta})+\abs{\hat\theta^1_t}^2+\abs{\hat\theta^2_t}^2+2(\hat\theta^1_t)^\top\rho_t\hat\theta^2_t\right]\,dt+\hat\theta^1_t\cdot dW_t+\hat\theta^2\cdot d\bar W_t\right.\\
	&\phantom{AA}\biggl.+\frac{\alpha_t}{Z_{t-}}\cdot dW_t+\Bigl[\frac{1}{\varphi(t,y)}-1\Bigr]\star N(dy,dt)+\lambda_t\Exp[\hat\varphi(t,Y_t)-1]\,dt\\
	&\phantom{AA}\biggl.+\frac{\bar{\alpha}_t}{Z_{t-}}\cdot d\bar{W}_t+\frac{\beta(t,y)}{Z_{t-}}\star \widetilde{N}(dy,dt)+\frac{1}{Z_{t-}}\Bigl[\left(\alpha_t\cdot\hat\theta_t^1+\bar{\alpha}_t\cdot\hat\theta_t^2+(\hat\theta^1)^\top\rho_t\bar{\alpha}_t+\alpha_t^\top\rho_t\hat\theta_t^2\right)dt \Bigr.\\
	&\phantom{AA}\Bigl.\Bigl.+\beta(t,y)\Bigl[\frac{1}{\hat\varphi(t,y)}-1\Bigr]\star N(dy,dt) \Bigr]\Bigr\}-D^{x,\hat\theta,\hat\varphi}_tdt.
	\end{align*}
Using (\ref{opt-cond}) and $\tilde N(dy,dt)=N(dy,dt)-F_t(dy)\lambda_t\,dt$ we obtain
\begin{align*}
&d\Bigl(\frac{Z_t}{H_t}\Bigr)=\frac{Z_{t-}}{H_{t-}}\left\{\left[r_t+\tilde{\vartheta}_t(\zeta^\theta)+\abs{\theta^1_t}^2+\abs{\theta^2_t}^2+2(\theta^1_t)^\top\rho_t\theta^2_t\right]\,dt+\theta^1_t\cdot dW_t+\theta^2\cdot d\bar W_t\right.\\
&\phantom{AAAA}\biggl.+\frac{\alpha_t}{Z_{t-}}\cdot dW_t+\frac{\bar{\alpha}_t}{Z_{t-}}\cdot d\bar{W}_t-\hat\kappa_t\cdot y\star {N}(dy,dt)+\lambda_t\Exp[\varphi(t,Y_t)\hat\kappa_t\cdot Y_t]\,dt\\
&\phantom{AAAA}\biggl.+\frac{1}{Z_{t-}}\left[\alpha_t\cdot\theta_t^1+\bar{\alpha}_t\cdot\theta_t^2+(\theta^1)^\top\rho_t\bar{\alpha}_t+\alpha_t^\top\rho_t \theta_t^2\right]\,dt \Bigr\}-D^{x,\hat\theta,\hat\varphi}_tdt.
\end{align*}
Dot products of (\ref{optcond-pi}) with $\hat\theta_t^1$ and $\hat\theta_t^2$ respectively, together with \eqref{optcond-hatfpi}, allow to transform the $dt$ term into
\begin{align*}
&r_t+\vartheta_t(\hpi_t)-\hpi_t\cdot[r_t\underline 1-\mu_t+\sigma_t(\hat\theta^1_t+\rho_t\hat\theta^2_t)]+\hpi_t^\top\sigma_t\hat\theta_t^1-\hat\kappa_t^\top b_t\hat\theta^2_t\\
&\phantom{AA}+\lambda_t\Exp[\varphi(t,Y_t)\hat\kappa_t\cdot Y_t]+\hpi_t^\top\sigma_t\rho_t\hat\theta_t^2-\hat\kappa_t^\top b_t\rho_t^\top\hat\theta_t^1
\end{align*}
which in turn equals $r_t\vartheta_t(\hpi_t)+\hpi_t\cdot(\mu_t-r_t\underline 1)+\hat\kappa_t\cdot p_t$ by condition \eqref{SDFcond1}. Using once again (\ref{optcond-pi}) we see that $Z^{\hat\theta,\hat\varphi}/H^{\hat\theta,\hat\varphi}$ solves the wealth equation \eqref{eqVnon-linear0} controlled by $(\hat\pi,\hat\kappa)$ and $D^{x,\hat\theta,\hat\varphi}.$
Now, since
\[
Z^{\hat\theta,\hat\varphi}_0/H^{\hat\theta,\hat\varphi}_0=\Exp[H_t^{\hat\theta,\hat\varphi}G_T^{x,\hat\theta,\hat\varphi}]=\calX^{\hat\theta,\hat\varphi}(\calY^{\hat\theta,\hat\varphi}(x))=x,
\]
by uniqueness of solution to \eqref{eqVnon-linear0} we obtain $V^{x,\hpi,\hat\kappa,\hat D}=Z^{\hat\theta,\hat\varphi}/H^{\hat\theta,\hat\varphi}.$ This implies, in particular, that $J(x;\hat\pi,\hat\kappa,D^{x,{\hat\theta,\hat\varphi}})=\bar J(G^{x,{\hat\theta,\hat\varphi}},D^{x,{\hat\theta,\hat\varphi}}).$

The proof that the strategy $\hat\pi,\hat\kappa,D^{x,{\hat\theta,\hat\varphi}}$ is admissible is the same as the proof of part (ii) of Lemma 1 in Michelbrik and Le \cite{michelbrink}. We can conclude that the strategy $(\hat\pi,\hat\kappa,\hat D)$ with $\hat D=D^{x,{\theta,\varphi}}$ is optimal.
\end{proof}

\begin{proof}[Proof of Proposition of \ref{prop_CRRA}] 
Let $({\theta,\varphi})\in\Theta$ and $x>0$ be fixed. For simplicity, as before we denote $H=H^{\theta,\varphi},$ $G=G^{x,\theta,\varphi}$ and $D=D^{x,\theta,\varphi}.$ Let $M$ be the martingale defined as
	\begin{equation*}
	M^{\theta,\varphi}_t:=\mathbb E\left[\Bigl.H_TG+\int_0^TH_sD_s\,ds\Bigr|\,\mathcal{F}_t \right], \quad t\in[0,T].
	\end{equation*}
	Notice this process satisfies
	\begin{equation*}
	M_t=Z_t+\int_0^tH_sD_s\,ds
	\end{equation*}
	for all $t\in[0,T]$, with $Z=Z_t^{\theta,\varphi}$ as in (\ref{linear-BSDE}). Hence, the processes $\alpha=\alpha^{\theta,\varphi}$, $\bar\alpha=\bar\alpha^{\theta,\varphi}$ and $\beta=\beta^{\theta,\varphi}$ are just the integrands in the martingale representation of $M$ with respect to $W$, $\bar W$ and $\widetilde{N}(dy,dt)$ respectively.
For CRRA preferences we have $I_1(t,y)=I_2(y)=y^{-\frac{1}{\eta}}$. Then $D_t=I_1(t,\calY(x)H_t)=[\calY(x)H_t]^{-1/\eta}$ and
\[
	\mathcal{X}(y)=\mathcal{X}^{\theta,\varphi}(y)=y^{-\frac{1}{\eta}}\mathcal{X}^{\theta,\varphi}(1), \ \ 
\mathcal Y(x)=\mathcal{Y}^{\theta,\varphi}(x)=\left[\frac{x}{\mathcal{X}(1)}\right]^{-\eta}
\]
with
	\begin{equation*}
	\mathcal{X}(1)=\mathbb E\left[\int_0^T H_t^{-\frac{1}{\eta}+1}\,dt+H_T^{-\frac{1}{\eta}+1} \right].
	\end{equation*}
	Hence, $D_t=\frac{x}{\mathcal X(x)}H_t^{-1/\eta}.$ Now, the process $H$ satisfies $H^{-\frac{1}{\eta}+1}=Lh$ with $L$ the exponential martingale solution to the linear SDE
	\begin{equation*}
	dL_t=L_{t-}\left\{\frac{1-\eta}{\eta}\left(\theta^1_t\cdot dW_t+\theta^2_t\cdot d\bar W_t\right)+ \left[\varphi(t,y)^{-\frac{1}{\eta}+1}-1\right]\star\widetilde{N}(dy,dt)\right\}
	\end{equation*}
	with $L_0=1,$ and $h$ deterministic, since it is the exponential of deterministic (Lebesgue) integrals of functions that depend only on $r_t,\lambda_t,F_t$ and $(\theta,\varphi).$ Then, $Z$ satisfies
\[	Z_t=\mathbb E\left[\Bigl. H_TG+\int_0^TH_sD_s\,ds
\,\Bigr|\,\mathcal{F}_t\right]\\
=\frac{x}{\mathcal{X}^{\theta,\varphi}(1)}L_t\left[h_T+\int_t^Th_s\,ds \right]\]
	for all $t\in[0,T]$. Using that $HD=\frac{x}{\mathcal X(x)}Lh,$ we obtain
	\begin{align*}
	dM_t&=dZ_t+H_tD_t\,dt\\
	&=\frac{x}{\mathcal{X}(1)}\left[h_T+\int_t^Th_sds \right]dL_t\\
	&=Z_{t-}\left\{\frac{1-\eta}{\eta}\left(\theta^1_t\cdot dW_t+\theta^2_t\cdot d\bar W_t\right)+\left[\varphi(t,y)^{-\frac{1}{\eta}+1}-1\right]\star\widetilde{N}(dy,dt)\right\}.
	\end{align*}
	The desired assertion follows by comparing coefficients of the last differential with those of the linear backward SDE (\ref{linear-BSDE}).
\end{proof}

\bibliographystyle{plain}
\bibliography{biblioALM}

\end{document}